\documentclass[12pt]{article}%
\usepackage{amsmath}
\usepackage{amsfonts}
\usepackage{amssymb}
\usepackage{graphicx}%
\providecommand{\U}[1]{\protect\rule{.1in}{.1in}}
\begin{document}

\title{Resonance structure of $^{8}$Be with the two-cluster resonating group method}
\author{ V. O. Kurmangaliyeva$^{1}$,  N.~Kalzhigitov$^{1}$, %\\
N. Zh. Takibayev$^{1}$,\\ V. S. Vasilevsky$^{2}$\\$^{1}$Al-Farabi Kazakh National University, Almaty, Kazakhstan\\$^{2}$Bogolyubov Institute for Theoretical Physics, Kiev, Ukraine}
\maketitle

\begin{abstract}
A two-cluster microscopic model is applied to study elastic alpha-alpha
 scattering and resonance structure of $^{8}$Be. The model is an algebraic
version of the Resonating Group Method, which makes use complete set of
oscillator functions to expand wave function of two-cluster system.
Interaction between clusters is determined by well-known semi-realistic
nucleon-nucleon potentials of Hasegawa-Nagata, Minnesota and Volkov. Detail
analysis of resonance  wave functions is carried out in oscillator,
coordinate and momentum spaces. Effects of the Pauli principle on wave
functions of the $^{8}$Be continuos spectrum states are thoroughly studied.

\end{abstract}

\section{Introduction}

Nucleus $^{8}$Be is a very interesting object which attracts large attention
from various theoretical and experimental methods. This nucleus has no bound
states but it has a reach collection of resonance states. The low-energy part
of resonance states belongs to the two-cluster continuum of two interacting
alpha particles. There are three resonance states $0^{+}$, $2^{+}$ and $4^{+}%
$, which are considered as a rotational band of the $^{8}$Be nucleus. The
$0^{+}$ resonance state is a very narrow resonance state with the energy 0.092
MeV above the $\alpha+\alpha$ decay threshold. The width of this state is 5.57
eV and the half -life time is 8.19$\times$10$^{-17}$seconds
\cite{2004NuPhA.745..155T}. This allows one to treat it approximately as a
bound state. As the threshold energy of the second channel $p+^{7}$Li is 17.35
MeV with respect to the energy of the first channel, then the $\alpha+\alpha$
channel is dominant in a wide energy range (0$\leq E\leq15$ MeV).

In the present paper we are going to study structure of $^{8}$Be  and pay
main attention to wave functions of resonance states. For this aim we employ a
specific variant (version) of the resonating group method (RGM) which was formulated
in Refs. \cite{kn:Fil_Okhr, kn:Fil81} and is known as the algebraic
version of the resonating group method. The key element of the algebraic
version is that it uses wave functions of three-dimension harmonic oscillator
to expand wave functions of inter-cluster motion and thus it realizes a matrix
form of quantum mechanical description of nuclear system. The matrix form can
be applied to study both bound and resonance states since corresponding
boundary conditions well-known in coordinate space were transformed into
discrete, oscillator representation. At present time, the algebraic version of
the resonating group method is very efficient and popular tool to study
dynamics of two-cluster systems (\cite{2015NuPhA.941..121L,
2017JPhCS.863a2015S, 2014JPhCS.569a2020S, 2ClBook2017,
 DUISENBAY2019Bul}), and three-cluster systems as well
(\cite{2010PPN....41..716N, PhysRevC.98.024325,
2017UkrJPh..62..461V, 2020NuPhA.99621692D}).

The present paper is a continuation of  the investigations of $^{8}$Be
started in Ref. \cite{2ClBook2017}. In the present paper we will reveal new
interesting information on peculiarities of dynamics of two-cluster systems
and their manifestations in $^{8}$Be. We discuss different ways of detecting
resonance states and we show how resonance states affect behavior of a
numerous number of observable (partial and total cross section of elastic
scattering and so on) and non-observable (wave function, an average distance
between clusters and so on) physical quantities. As our model involves
oscillator basis, we then analyze behavior of wave functions in oscillator,
coordinate and momentum representations. An explicit evidence of a simple
relation between expansion coefficients and corresponding wave function in
coordinate space will be shown for a continuous spectrum state.

The layout of the present paper is the following. In Sec. \ref{Sec:Method} we
shortly review principal ideas of the two-cluster model. In this section we
also introduce all necessary quantities which will be used to analyze obtained
results. Sec. \ref{Sec:Analysis} will start with selecting of nucleon-nucleon
potentials and with fixing all input parameters of our calculations. Then it
proceeds with analysis of phase shifts of elastic alpha-alpha scattering and
determination of the energy and width of resonance states. After that,
peculiarities of wave functions of resonance states and effects of the Pauli
principle on continuous spectrum states of $^{8}$Be are discussed in detail.
The final section summarizes main results of our investigations.

\section{Method \label{Sec:Method}}

Formulation of a microscopic method requires to display many-particle
Hamiltonian and an explicit form of wave function. Hamiltonian which will be
used in our calculations involves the kinetic energy operator, a
semi-realistic nucleon-nucleon potential and the Coulomb interaction of
protons. To achieve our goals, the sought wave function is selected to
reproduce two-cluster structure of $^{8}$Be. Therefore, wave function of
$^{8}$Be comprised of two alpha particles is represented in the form%
\begin{equation}
\Psi_{EL}=\widehat{\mathcal{A}}\left\{  \Phi_{1}\left(  ^{4}\text{He}%
,b\right)  \Phi_{2}\left(  ^{4}\text{He},b\right)  \psi_{EL}\left(  q\right)
Y_{LM}\left(  \widehat{\mathbf{q}}\right)  \right\}  , \label{eq:M10}%
\end{equation}
where $\widehat{\mathcal{A}}$ is the antisymmetrization operator, $\Phi
_{1}\left(  ^{4}\text{He},b\right)  $ and $\Phi_{2}\left(  ^{4}\text{He}%
,b\right)  $ are the translational invariant and antisymmetric functions
describing internal structure of the first and second alpha clusters,
respectively. Since spin of an alpha particle equals zero, then the total spin
$S$ of $^{8}$Be equals also zero and therefore the total angular momentum $J$
coincides with the total orbital momentum $L$. Throughout of the text we will
use the total orbital momentum $L$ to mark different rotation states of $^{8}%
$Be. A wave function $\psi_{EL}\left(  q\right)  $ represents radial motion
two clusters, while the spherical harmonic $Y_{LM}\left(  \widehat{\mathbf{q}%
}\right)  $ represents rotating motion of clusters. The Jacobi vector
$\mathbf{q}=q\cdot\widehat{\mathbf{q}}$ ($\widehat{\mathbf{q}}$ is a unit
vector orientation of which in space is fixed by two angles $\theta_{q}$\ and
$\phi_{q}$) is proportional to the distance $\mathbf{r}$ between interacting
clusters%
\begin{equation}
\mathbf{q}=\mathbf{r}\sqrt{\frac{A_{1}A_{2}}{A_{1}+A_{2}}}=\sqrt{\frac
{A_{1}A_{2}}{A_{1}+A_{2}}}\left[  \frac{1}{A_{1}}\sum_{i\in A_{1}}%
\mathbf{r}_{i}-\frac{1}{A_{2}}\sum_{j\in A_{2}}\mathbf{r}_{j}\right]  ,
\label{eq:M11}%
\end{equation}
where $\mathbf{r}_{1}$, $\mathbf{r}_{2}$, \ldots, $\mathbf{r}_{A}$ are
coordinates in the space of individual nucleons.

The main assumption of the RGM is that wave functions $\Phi_{1}\left(
^{4}\text{He},b\right)  $ and $\Phi_{2}\left(  ^{4}\text{He},b\right)  $ are
known and fixed, while the inter-cluster function $\psi_{EL}\left(  q\right)
$ has to be obtained by solving the dynamic equations. In the standard version
of the RGM, one has to solve the integro-differential equation. The integral
or nonlocal part of the equation appears due to the antisymmetrization
operator or, in other words, due to the Pauli principle. In the algebraic
version of RGM, the dynamic equations transforms in to \ a set of linear
algebraic equations. This is achieved by using a full set of the radial part
of oscillator functions $\left\{  \Phi_{nL}\left(  q,b\right)  \right\}  $. By
expanding the inter-cluster function $\psi_{EL}\left(  q\right)  $ over
oscillator functions%

\begin{equation}
\psi_{EL}\left(  q\right)  =\sum_{n=0}^{\infty}C_{nL}^{\left(  E\right)  }%
\Phi_{nL}\left(  q,b\right)  \label{eq:M12}%
\end{equation}
or the total two-cluster function $\Psi_{EL}$ over cluster oscillator
functions $\left\{  \left\vert nL\right\rangle \right\}  $%

\begin{equation}
\Psi_{EL}=\sum_{n=0}^{\infty}C_{nL}^{\left(  E\right)  }\left\vert
nL\right\rangle , \label{eq:M12A}%
\end{equation}
we arrive to a system of linear algebraic equations%
\begin{equation}
\sum_{\widetilde{n}=0}^{\infty}\left\{  \left\langle nL\left\vert \widehat
{H}\right\vert \widetilde{n}L\right\rangle -E\delta_{n\widetilde{n}}%
\Lambda_{nL}\right\}  C_{\widetilde{n}L}^{\left(  E\right)  }=0.
\label{eq:M13}%
\end{equation}
The cluster oscillator function is determined as%
\begin{equation}
\left\vert nL\right\rangle =\widehat{\mathcal{A}}\left\{  \Phi_{1}\left(
^{4}\text{He},b\right)  \Phi_{2}\left(  ^{4}\text{He},b\right)  \Phi
_{nL}\left(  q,b\right)  Y_{LM}\left(  \widehat{\mathbf{q}}\right)  \right\}
. \label{eq:M14}%
\end{equation}
One can see that an oscillator function $\Phi_{nL}\left(  q,b\right)  $
describes relative motion of two alpha particles in the cluster function
(\ref{eq:M14}). And here is the explicit form of the oscillator function
\begin{eqnarray}
\Phi_{nL}\left(  q,b\right)   &  =& \left(  -1\right)  ^{n}\mathcal{N}%
_{nL}~b^{-3/2}\rho^{L}e^{-\frac{1}{2}\rho^{2}}L_{n}^{L+1/2}\left(  \rho
^{2}\right)  ,\quad\label{eq:M17}\\
\rho &  =& q/b.\nonumber
\end{eqnarray}
As we interested in the inter-cluster wave function in the momentum space
$\psi_{EL}\left(  p\right)  $, we present also oscillator functions in
momentum space%
\begin{eqnarray}
\Phi_{nL}\left(  p,b\right)   &  = & \mathcal{N}_{nL}~b^{3/2}\rho^{L}e^{-\frac
{1}{2}\rho^{2}}L_{n}^{L+1/2}\left(  \rho^{2}\right)  ,\label{eq:M18}\\
\quad\rho &  = &p\cdot b,\nonumber
\end{eqnarray}
where%
\[
\mathcal{N}_{nL}=\sqrt{\frac{2\Gamma\left(  n+1\right)  }{\Gamma\left(
n+L+3/2\right)  }}%
\]
and $L_{n}^{\alpha}\left(  z\right)  $ is the generalized Laguerre polynomial
\cite{kn:abra}.

The system of equations (\ref{eq:M13}) contains matrix elements of Hamiltonian
between cluster oscillator functions $\left\langle nL\left\vert \widehat
{H}\right\vert \widetilde{n}L\right\rangle $ and matrix elements of unit
operator $\left\langle nL|\widetilde{n}L\right\rangle $. For two-cluster
systems, matrix elements are diagonal with respect to quantum numbers $n$ and
$\widetilde{n}$ \ and coincides with the so-called eigenvalues of the norm
kernel $\Lambda_{nL}$%
\[
\left\langle nL|\widetilde{n}L\right\rangle =\delta_{n,\widetilde{n}}%
\Lambda_{nL}.
\]
For $^{8}$Be the eigenvalues $\Lambda_{nL}$ equal%
\begin{equation}
\Lambda_{nL}=\frac{1}{2}\sum_{k=0}^{4}\frac{4!\left(  -1\right)  ^{k}%
}{k!\left(  4-k\right)  !}\left[  1-k\frac{1}{2}\right]  ^{2n+L}.
\label{eq:A1}%
\end{equation}

The system of equations (\ref{eq:M13}) is deduced directly from the
Schr\"{o}dinger equation
\[
\left(  \widehat{H}-E\right)  \Psi_{EL}=0
\]
for the wave function (\ref{eq:M10}).

By solving the set of equations (\ref{eq:M13}), one obtains the energy and a
wave function of bound states, or a wave function and the scattering
$S$-matrix for continuous spectrum states. If in Eq. (\ref{eq:M13}) we
restrict ourselves with a finite number (we denote it $N$) of oscillator
function ($n$=0, 1, \ldots, $N-1$), \ we encounter the generalized eigenvalue
problem for $N\times N$ matrices. By solving this problem, we obtain the
energy spectrum $E_{\nu}$ ($\nu$=1, 2, \ldots, $N$) and wave functions
$\left\{  C_{nL}^{\left(  E_{\nu}\right)  }\right\}  $ of bound and
pseudo-bound states. The latter are continuous spectrum states describing
alpha-alpha scattering \ states with specific conditions. It was shown in Ref.
\cite{kn:VVS+1983Li7E}\ that the wave functions of pseudo-bound states in
oscillator space has a node at the point $N$, i.e.
\[
C_{NL}^{\left(  E_{\nu}\right)  }=0.
\]
Thus, diagonalization of Hamiltonian with a fixed number of oscillator
functions selects from continuous spectrum states those states which obey
specific boundary condition.

To solve the system of equations (\ref{eq:M10}) for a scattering state, one
has to formulate proper boundary conditions in\ discrete oscillator space and
then incorporate them in a set of equations (\ref{eq:M13}).\textit{ }This
problem has been numerously discussed in literature (see, for instance,
\cite{kn:Yamani}, \cite{kn:Heller1} \cite{kn:Fil_Okhr}, \cite{kn:Fil81},
\cite{kn:cohstate2E}). Here we shortly outline practical steps to obtain and
analyze scattering states.

In oscillator space like in coordinate space, we split the space on two parts
or two regions: internal and asymptotic regions. Let us recall how boundary
conditions are formulated and used in coordinate space. In the internal region
interaction between clusters are prominent and it should be treated correctly.
In the asymptotic region interaction between clusters originated from a
short-range nucleon-nucleon potential is negligibly small, and can therefore
be ignored. In consequence of this fact, in the asymptotic region Hamiltonian
consists of the kinetic energy operator $\widehat{T}_{q}$ of relative motion
of clusters for neutral clusters (or when one of the clusters is a neutron),
for charged clusters it consists of the same kinetic energy operator and the
Coulomb interaction of two point-like charged particles
\begin{equation}
\widehat{H}=\widehat{T}_{q}+\frac{Z_{1}Z_{2}e^{2}}{q}\sqrt{\mu},\label{eq:C10}%
\end{equation}
Where $Z_{1}$($Z_{2}$)\ is a charge of the first (second) cluster,
$e^{2}=1.44$ MeV$\cdot$fm is the square of the elementary charge in nuclear
units, and
\begin{eqnarray}
\widehat{T}_{q} &  = & -\frac{\hbar^{2}}{2m}\left[  \frac{d^{2}}{dq^{2}}%
+2\frac{1}{q}\frac{d}{dq}-\frac{L\left(  L+1\right)  }{q^{2}}\right]
,\label{eq:C10A}\\
\mu &  =& \frac{A_{1}A_{2}}{A_{1}+A_{2}}.\nonumber
\end{eqnarray}

It is well-known (see, for example, books \cite{BazBookE}, \cite{kn:Newton})
that Hamiltonian (\ref{eq:C10}) has two independent solutions
\begin{equation}
\psi_{kL}^{\left(  R\right)  }\left(  q\right)  =\sqrt{\frac{2}{\pi}}%
k\frac{F_{L}\left(  \rho,\eta\right)  }{\rho},\quad\psi_{kL}^{\left(
I\right)  }\left(  q\right)  =\sqrt{\frac{2}{\pi}}k\frac{G_{L}\left(
\rho,\eta\right)  }{\rho}, \label{eq:C11}%
\end{equation}
where $\rho=kq$, $\eta$\ is the Sommerfeld parameter:%
\begin{equation}
\eta=\frac{Z_{1}Z_{2}e^{2}}{\hbar k}\sqrt{\mu m}. \label{eq:C11B}%
\end{equation}
Wave functions $\psi_{kL}^{\left(  R\right)  }\left(  q\right)  $ and
$\psi_{kL}^{\left(  I\right)  }\left(  q\right)  $ describes scattering of
charged particles and are regular and irregular (singular) at the origin of
coordinates. For neutral particles, when $\eta=$0, these functions are the
spherical Bessel and Neumann functions%
\begin{equation}
\psi_{kL}^{\left(  R\right)  }\left(  q\right)  =\sqrt{\frac{2}{\pi}}%
kj_{L}\left(  \rho\right)  ,\quad\psi_{kL}^{\left(  I\right)  }\left(
q\right)  =-\sqrt{\frac{2}{\pi}}kn_{L}\left(  \rho,\eta\right)  ,
\label{eq:C11D}%
\end{equation}

General solutions of Hamiltonian (\ref{eq:C10})\ is the following combination
of regular and irregular functions%
\begin{equation}
\psi_{kL}^{\left(  a\right)  }\left(  q\right)  =\psi_{kL}^{\left(  R\right)
}\left(  q\right)  +\tan\delta_{L}\psi_{kL}^{\left(  I\right)  }\left(
q\right)  , \label{eq:C12}%
\end{equation}
where $\delta_{L}$ is a phase shift of elastic cluster-cluster scattering. If
we managed to calculate the phase shift $\delta_{L}$, then we immediately
determine two-cluster wave function in semi-infinite range of distances
$R_{i}\leq q<\infty$, where $R_{i}$ indicates the inter-cluster distance which
marks a border between the internal and asymptotic regions. Eq. (\ref{eq:C12})
explicitly demonstrate boundary conditions for scattering states in a
single-channel case. The asymptotic wave function $\psi_{kL}^{\left(
a\right)  }\left(  q\right)  $ at the point $q=R_{i}$ has to be matched with
the internal wave function $\psi_{kL}^{\left(  i\right)  }\left(  q\right)  $,
which we assume is determined by numerical solution of the Schr\"{o}dinger
equation in the range of inter-cluster distances $0\leq q<R_{i}$. The boundary
conditions then are read as%
\begin{eqnarray}
\psi_{kL}^{\left(  i\right)  }\left(  R_{i}\right)   &  = & \psi_{kL}^{\left(
a\right)  }\left(  R_{i}\right)  ,\label{eq:C13A}\\
\left.  \frac{d}{dq}\psi_{kL}^{\left(  i\right)  }\left(  q\right)
\right\vert _{q=R_{i}}  &  =& \left.  \frac{d}{dq}\psi_{kL}^{\left(  a\right)
}\left(  q\right)  \right\vert _{q=R_{i}}, \label{eq:C13B}%
\end{eqnarray}
where the asymptotic and internal wave functions are matched and their first
derivatives as well. These conditions guarantee that the inter-cluster wave
function and its first derivative are continuos at the point $q=R_{i}$.

Essentially the same ideas were used to formulate boundary in oscillator
representations. Here we present the shortest way of explanation but not
completely rigorous. For the sake of simplicity, we consider neutral clusters.
As we pointed out above, Hamiltonian in asymptotic region consists of the
kinetic energy operator. In oscillator representation, this Hamiltonian has a
tridiagonal or Jacobi matrix form. Nonzero matrix elements $\left\langle
m,L\left\vert \widehat{T}_{q}\right\vert n,L\right\rangle $ of the kinetic
energy operator are%
\begin{equation}
\left\langle mL\left\vert \widehat{T}_{q}\right\vert nL\right\rangle
=\frac{\hbar^{2}}{2mb^{2}}\left\{
\begin{array}
[c]{cc}%
-\sqrt{n\left(  n+L+\frac{1}{2}\right)  } & m=n-1\\
\left(  2n+L+\frac{3}{2}\right)  & m=n\\
-\sqrt{\left(  n+1\right)  \left(  n+L+\frac{3}{2}\right)  } & m=n+1
\end{array}
\right.  . \label{eq:C22}%
\end{equation}

It was shown in Refs. \cite{kn:Heller1}, \cite{1975JMP....16..410Y},
\cite{kn:SmirnovE} that the matrix equation%
\begin{equation}
\sum_{n=0}^{\infty}\left[  \left\langle mL\left\vert \widehat{T}%
_{q}\right\vert nL\right\rangle -E\delta_{mn}\right]  C_{nL}=0 \label{eq:C22A}%
\end{equation}
has two independent solutions $C_{nL}^{\left(  R\right)  }\left(  kb\right)  $
and $C_{nL}^{\left(  I\right)  }\left(  kb\right)  $, traditionally we refer
to them as to regular and irregular solutions of the equations (\ref{eq:C22A}%
). Explicit form of the solutions $C_{nL}^{\left(  R\right)  }\left(
kb\right)  $ and $C_{nL}^{\left(  I\right)  }\left(  kb\right)  $ can be found
in Ref. \cite{kn:Heller1}, \cite{1975JMP....16..410Y}, \cite{kn:SmirnovE}. It
was shown in Refs. \cite{kn:Fil_Okhr}, \cite{kn:Fil81} \ that for large values
of $n$ the expansion coefficients $C_{nL}^{\left(  R\right)  }\left(
kb\right)  $ and $C_{nL}^{\left(  I\right)  }\left(  kb\right)  $ are
connected to the regular and irregular functions (\ref{eq:C11D}) by the simple
relations%
\begin{equation}
C_{nL}^{\left(  R\right)  }\left(  kb\right)  \approx\sqrt{2R_{n}}\psi
_{kL}^{\left(  R\right)  }\left(  R_{n}\right)  ,\quad C_{nL}^{\left(
I\right)  }\left(  kb\right)  \approx\sqrt{2R_{n}}\psi_{kL}^{\left(  I\right)
}\left(  R_{n}\right)  , \label{eq:C23}%
\end{equation}
where
\begin{equation}
R_{n}=b\sqrt{4n+L+3} \label{eq:C23A}%
\end{equation}
\ is the turning point of the classical harmonic oscillator. The relations
(\ref{eq:C23}) reflect \ general properties of oscillator functions. To this
end, relations similar to (\ref{eq:C23}) are valid for any functions of a
two-cluster system and corresponding expansion coefficients. It was explicitly
demonstrated in Ref. \cite{Kalzhigitov2020Bul} \ for the wave function of the
$^{6}$Li ground state. It was shown that the relation between wave function
and expansion coefficients (\ref{eq:C23}) is valid even for small values of
$n$ ($n\geq5$). \ In present calculations we use similar relations for the
expansion coefficients of the asymptotic wave functions (\ref{eq:C11}) for
charged clusters.

To solve Eq. (\ref{eq:M13}) for continuous spectrum states, we have to
introduce appropriate boundary conditions in that set of equations. For this
aim an infinite set of expansion coefficients is divided on two subsets -
internal and asymptotic:%
\begin{equation}
\left\{  C_{nL}^{\left(  E\right)  }\right\}  =\left\{  C_{0L}^{\left(
E\right)  },C_{1L}^{\left(  E\right)  },\ldots,C_{N-1L}^{\left(  E\right)
},C_{\nu L}^{\left(  R\right)  }+\tan\delta_{L}C_{\nu L}^{\left(  I\right)
}\right\}  , \label{eq:C24}%
\end{equation}
where index $\nu$ ($N\leq\nu<\infty$) numerates expansion coefficients in the
asymptotic region. With such form of expansion coefficients we have got $N+1$
unknown quantities ($N$ expansion coefficients in internal region and phase
shift $\delta_{L}$) to be determined. By substituting the expansion
coefficients (\ref{eq:C24}) in Eq. (\ref{eq:M13}), we obtain%
\begin{eqnarray}
& & \sum_{\widetilde{n}=0}^{N-1}\left\{  \left\langle nL\left\vert \widehat
{H}\right\vert \widetilde{n}L\right\rangle -E\left\langle nL|\widetilde
{n}L\right\rangle \right\}  C_{nL}^{\left(  E\right)  }+\tan\delta_{L}%
\sum_{v\geq N}\left\langle nL\left\vert \widehat{H}\right\vert \nu
L\right\rangle C_{\nu L}^{\left(  I\right)  }\label{eq:C25}\\
& & =-\sum_{v\geq N}\left\langle nL\left\vert \widehat{H}\right\vert \nu
L\right\rangle C_{\nu L}^{\left(  R\right)  }.\nonumber
\end{eqnarray}
It is important to underline that in this system of equations, index $n$ is
run from 0 to $N-1$, and thus we have a set of $N+1$ linear algebraic
equations for $N+1$ unknown quantities. \ By taking into account Eq.
(\ref{eq:C22A}), we can simply the set of equations (\ref{eq:C25})%
\begin{eqnarray}
& & \sum_{\widetilde{n}=0}^{N-1}\left\{  \left\langle nL\left\vert \widehat
{H}\right\vert \widetilde{n}L\right\rangle -E\left\langle nL|\widetilde
{n}L\right\rangle \right\}  C_{nL}^{\left(  E\right)  }+\tan\delta_{L}%
\sum_{v\geq N}\left\langle nL\left\vert \widehat{V}\right\vert \nu
L\right\rangle C_{\nu L}^{\left(  I\right)  }\label{eq:C25A}\\
& & =-\sum_{v\geq N}\left\langle nL\left\vert \widehat{V}\right\vert \nu
L\right\rangle C_{\nu L}^{\left(  R\right)  }.\nonumber
\end{eqnarray}
This is basic set of equations which allows us to obtain a phase shift and
wave function of continuous spectrum states. Numerical solution of this set of
equations is performed with a finite sum of index $\nu$, in our calculations
presented bellow the sum involves 20 terms.

We will not also dwell on calculating of matrix elements of the kinetic and
potential energy operators, as their explicit form and reliable methods of
their calculations can be found in Ref. \cite{kn:cohstate2E}.

It is important to note \ that wave function $\Psi_{EL}$ for bound and
pseudo-bound states is traditionally normalized to unit%
\[
\left\langle \Psi_{EL}|\Psi_{EL}\right\rangle =\sum_{n=0}^{\infty}\left\vert
C_{nL}^{\left(  E\right)  }\right\vert ^{2}=1,
\]
however, corresponding inter-cluster function is normalized as
\begin{equation}
\left\langle \psi_{EL}|\psi_{EL}\right\rangle =S_{L}. \label{eq:C19}%
\end{equation}
In oscillator representation it can be represented as%
\begin{equation}
S_{L}=\sum_{n=0}^{\infty}\left\vert C_{nL}^{\left(  E\right)  }\right\vert
^{2}/\lambda_{n} \label{eq:C19A}%
\end{equation}
The quantity $S_{L}$ is proportional to the spectroscopic factor $SF_{L}$ (see
definition, for instance, in \cite{1960RvMP...32..567M}, \cite{Glendenning83}
and \cite{Nemec88E}, Chapter 9), which play an important role in the theory of
nuclear reactions when the Pauli principle is treated approximately
\cite{Glendenning83}. The factor $SF_{L}$ is used to determine amount of a
certain (definite) clusterization in a wave function of the compound system.
It is obvious from the definition of the spectroscopic factor (\ref{eq:C19}),
that it can be determined for bound state only, when norms of wave functions
$\Psi_{EL}$ and $\psi_{EL}$\ are finite.

\subsection{Definition of basic quantities}

Having calculated the wave function of the ground or resonance state in the
oscillator representation $C_{nL}^{\left(  E\right)  }$, we can easily
construct the inter-cluster wave functions in the coordinate and momentum
representations%
\begin{eqnarray}
\psi_{EL}\left(  q\right)   &  = & \sum_{n=0}C_{nL}^{\left(  E\right)  }\Phi
_{nL}\left(  q,b\right)  ,\label{eq:209A}\\
\psi_{EL}\left(  p\right)   &  =& \sum_{n=0}C_{nL}^{\left(  E\right)  }\Phi
_{nL}\left(  p,b\right)  .\label{eq:209B}%
\end{eqnarray}

One can use these functions to obtain additional information about bound and
resonance states. We will employ wave function of coordinate space $\psi
_{EL}\left(  q\right)  $ to calculate an average distance between clusters
$A_{c}$, which can be defined as%
\begin{equation}
A_{c}=b\sqrt{\left\langle \psi_{EL}\left(  q\right)  \left\vert q^{2}%
\right\vert \psi_{EL}\left(  q\right)  \right\rangle /\mu}, \label{eq:M19A}%
\end{equation}
where $\mu=A_{1}A_{2}/A$\ is the reduced mass for selected clusterization.
Similarly, we can also determine an average momentum $P_{c}$ of relative
motion of two clusters
\begin{equation}
P_{c}=b^{-1}\sqrt{\left\langle \psi_{EL}\left(  p\right)  \left\vert
p^{2}\right\vert \psi_{EL}\left(  p\right)  \right\rangle }. \label{eq:M19B}%
\end{equation}
The average momentum $P_{c}$ is related to an average kinetic energy of
two-cluster relative motion by simple relation%
\[
T_{c}=\frac{\hbar^{2}}{2m}P_{c}^{2}.
\]

By considering continuous spectrum states we will calculate and analyze the
weight of internal part of wave function, which we define as%
\begin{equation}
W_{L}\left(  E\right)  =\sum_{n=0}^{N_{i}-1}\left\vert C_{nL}^{\left(
E\right)  }\right\vert ^{2}.\label{eq:M20}%
\end{equation}
This definition of the weight equivalent to following definition%
\begin{equation}
W_{L}\left(  E\right)  =\int_{0}^{R}dqq^{2}\left\vert \psi_{EL}\left(
q\right)  \right\vert ^{2},\label{eq:M21}%
\end{equation}
the radius of internal region $R$ can be determined in self-consistent way as
\[
R\approx b\sqrt{4N_{i}+2L+3}.
\]
Let us evaluate behavior of function $W_{L}\left(  E\right)  $ for a simple
case when the function $\psi_{EL}\left(  q\right)  $ describes a free motion
of two cluster with the orbital momentum $L$=0. Then, the wave function
$\psi_{E}\left(  q\right)  =$ $\psi_{E,L=0}\left(  q\right)  $ is%
\[
\psi_{E}\left(  q\right)  =\sqrt{\frac{2}{\pi}}\frac{\sin\left(  kq\right)
}{q},
\]
and the weight $W\left(  E\right)  $ is equal to%
\[
W\left(  E\right)  =\int_{0}^{R}dqq^{2}\left\vert \psi_{E}\left(  q\right)
\right\vert ^{2}=\frac{2}{\pi}\left[  \frac{1}{2}R-\frac{\sin\left(
2kR\right)  }{4k}\right]  ,
\]
were $k$ is the wave number%
\[
k=\sqrt{\frac{2mE}{\hbar^{2}}}.
\]
Thus, in a simple case the weight $W\left(  E\right)  $ as a function of
energy has an oscillatory behavior and is decreased with increasing of energy
$E$ or wave number $k$. It is interesting to note that for small values of $E$
we have got
\[
W\left(  E\right)  \approx\frac{2}{3\pi}R\left(  kR\right)  ^{2},
\]
the weight $W\left(  E\right)  $ equal zero at $E=0$, and the slowly
increasing as a linear function of energy. Such a behavior of weights as a
function of energy suggests that this function has a maximum at relatively
small energy. We will see later that the weights $W\left(  E\right)  $ of
internal part of \ scattering wave functions allow us to find position of
resonance states and evaluate its width.

Within our method, parameters of resonance states are obtained from
corresponding phase shifts by using the Breit-Wigner formula for a phase shift
around a resonance state. It assumed that the phase shift $\delta$ in the
vicinity of a resonance state consists background (potential) phase shift
\ $\delta_{p}$ and resonance phase shift $\delta_{R}$
\[
\delta=\delta_{p}+\delta_{R},
\]
where the resonance phase shift is determined by the Breit-Wigner formula
\[
\delta_{R}=-\arctan\left(  \frac{\Gamma}{E-E_{r}}\right)  .
\]
We also assume that the first derivative of the background phase shift with
respect to energy is much more smaller than the first derivative of the
resonance phase shift with respect to energy. Then, the energy and width of
resonance state can be determined from the following equations:%
\begin{equation}
\left.  \frac{d^{2}\delta\left(  E\right)  }{dE^{2}}\right\vert _{E=E_{r}%
}=0,\quad\Gamma=2\left[  \frac{d\delta\left(  E\right)  }{dE}\right]
_{E=E_{r}}^{-1}. \label{eq:R01}%
\end{equation}
These relations allows us to determine more correctly the energy and width of
\ a resonance state especially in the case when the background phase shift is
not small and when the resonance state is wide.

\section{Results and discussion \label{Sec:Analysis}}

Before start doing calculations we need to select nucleon-nucleon potential
and to fix some parameters. To analyze structure of $^{8}$Be, we selected
three semi-realistic nucleon-nucleon potentials which are very often used in
many microscopical models of light atomic nuclei. They are the Volkov N2
potential (VP) \cite{kn:Volk65}, the Minnesota potential (MP)
\cite{kn:Minn_pot1}, \cite{1970NuPhA.158..529R} and modified Hasegawa-Nagata
potential (MHNP) \cite{potMHN1, potMHN2}. These potentials provide fairly good
description of nucleon-nucleon interactions in the states with different
values of two-nucleon spin $S$ and isospin $T$. They also give acceptable
energy and size parameters of light nuclei described by a wave function of
many-particle shell model. First of all, we have one free parameter in our
calculation. It is the oscillator length $b$. \ It is natural to choose such
value $b$ which minimize the binding energy of an alpha particle. We slightly
adjusted the Majorana exchange parameter $m$ in the VP and MHNP, and the
exchange parameter $u$ of the MP. The adjusted and original exchange
parameters are shown in Table \ref{Tab:NNparam}. The optimal value of the
oscillator length $b$ is also indicated in Table \ref{Tab:NNparam}.%

%TCIMACRO{\TeXButton{B}{\begin{table}[tbp] \centering}}%
%BeginExpansion
\begin{table}[tbp] \centering
%EndExpansion
\caption{Optimal values of oscillator length $b$ and exchange parameter 
$m$ or $u$ of nucleon-nucleon potentials.}%
\begin{tabular}
[c]{|c|c|c|c|}\hline
Potential & $b$, fm & $m/u$, adjusted & $m/u$, original\\\hline
VP & 1.376 & 0.6011 & 0.600\\
MP & 1.285 & 0.9347 & -\\
MHNP & 1.317 & 0.3961 & 0.4057\\\hline
\end{tabular}
\label{Tab:NNparam}%
%TCIMACRO{\TeXButton{E}{\end{table}}}%
%BeginExpansion
\end{table}%
%EndExpansion

\begin{figure}[ptb]
\begin{center}
\includegraphics[
%natheight=6.406500in,
%natwidth=7.506600in,
%height=11.2994cm,
width=13.2259cm
]{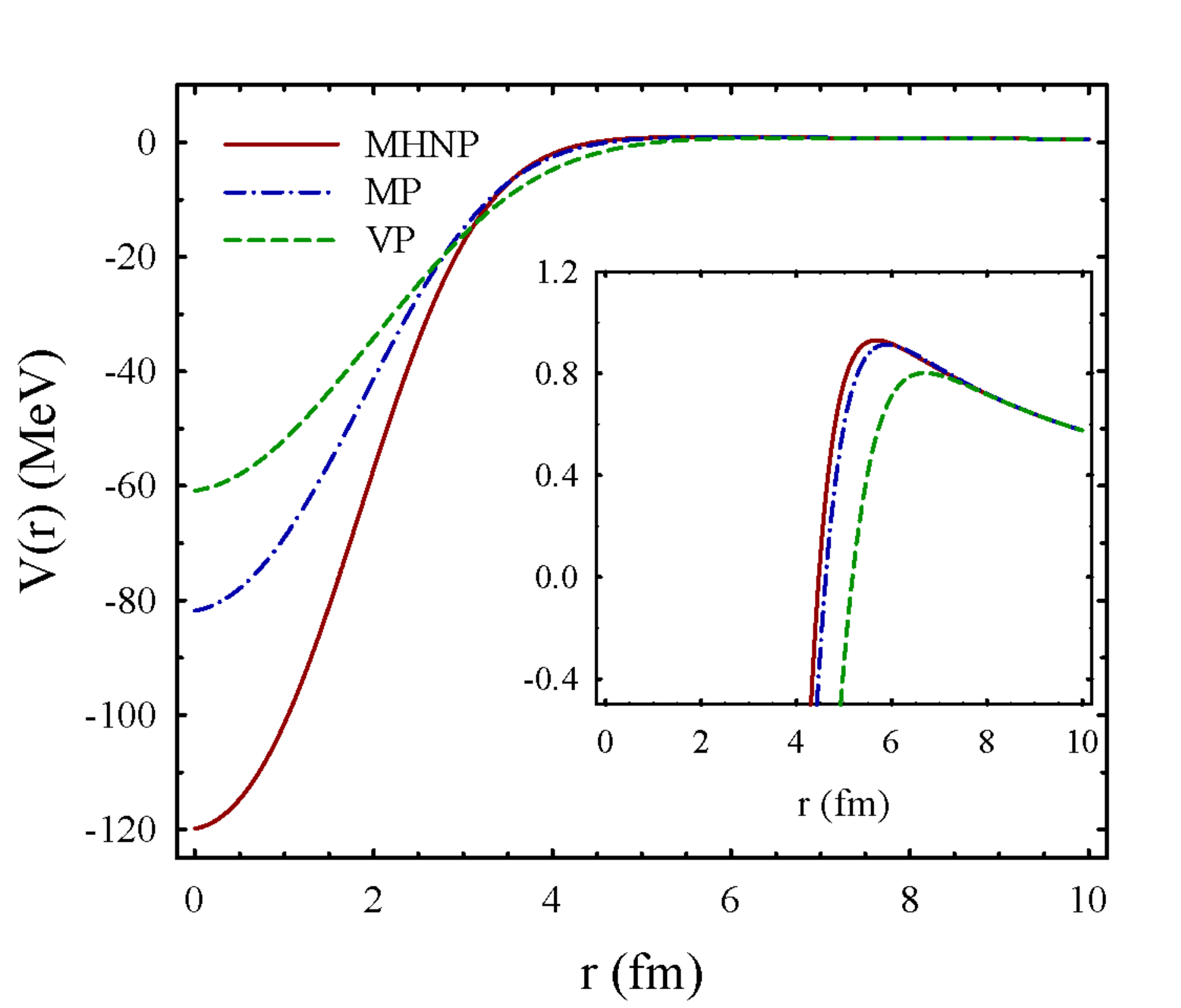}%
%{FoldPot8Be.jpg}%
\caption{Folding potential of the alpha-alpha interaction generated by the
MHNP, MP and VP.}%
\label{Fig:FoldPot8Be}%
\end{center}
\end{figure}

It is interesting to compare cluster-cluster potential generated by the
selected nucleon-nucleon potentials. Unfortunately, it is difficult task since
the cluster-cluster potential is nonlocal and energy-dependent interaction.
This is due to the Pauli principle. However, we can compare so-called a
folding potential which is local and represents the main part of
cluster-cluster interaction when the antisymmetrization between clusters is
disregarded. It means that in Eq. (\ref{eq:M10}) the antisymmetrization
operator $\widehat{\mathcal{A}}$ is set to be $\widehat{\mathcal{A}}=1$. In
Fig. \ref{Fig:FoldPot8Be} the folding potentials generated by three
nucleon-nucleon potentials are shown as a function of the inter-cluster
distance. The folding potentials like nucleon-nucleon potentials have the
Gaussian shape. However, contrary to nucleon-nucleon potentials, the folding
potentials have no repulsive core at small distances between alpha particles.
As the folding potential \ involves the Coulomb interaction of protons, there
is a barrier which stipulates existence of the $0^{+}$ resonance state. One
can see, that the MP\ and MHNP generates barrier of the same shape and
height.
%BeginExpansion
%EndExpansion

\begin{figure}
[ptb]
\begin{center}
\includegraphics[
%natheight=6.406500in,
%natwidth=7.373400in,
%height=11.462cm,
width=13.1797cm
]{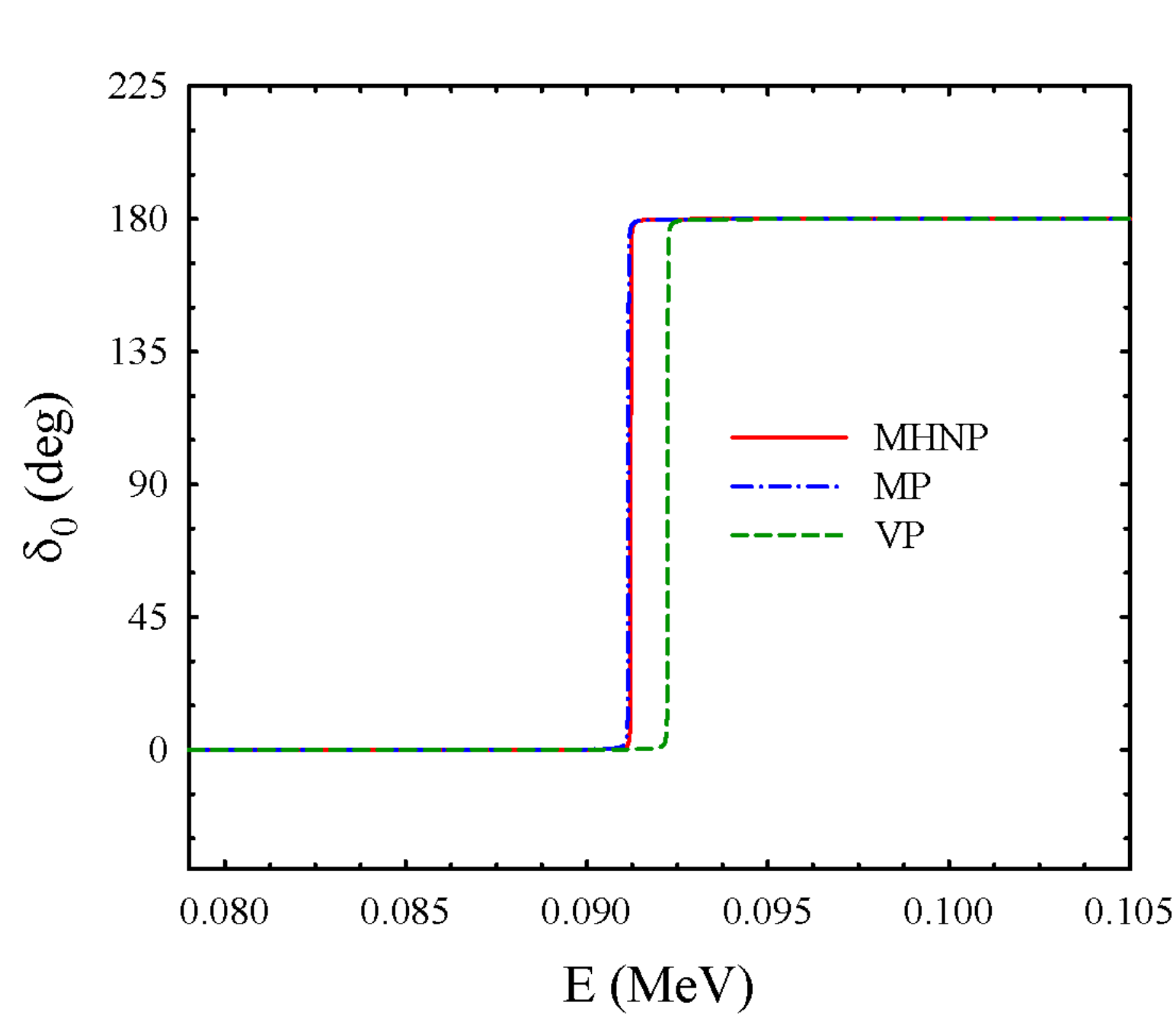}%
%{Phases8BeL0.jpg}%
\caption{Phase shifts of the alpha-alpha scattering with zero value of the total
orbital momentum.}%
\label{Fig:Phases8BeL0}%
\end{center}
\end{figure}

We start our investigations with the $s$ phase shift of the elastic
alpha-alpha scattering and position of the $0^{+}$ resonance state. In Fig.
\ref{Fig:Phases8BeL0} we show behavior of the $s$ phase shift around the
$0^{+}$ resonance state. As we can see this example of the classical
Breit-Wigner resonance state since phase shift is increased on 180$^{\circ}$
at the resonance energy and the background phase shift equals zero before and
\ after that energy. We manage to%

%BeginExpansion
%EndExpansion

\subsection{Spectrum of resonance states}

In Table \ref{Tab: Resonan8Be} we compare results of our calculations with the
available experimental data \cite{2004NuPhA.745..155T}. Energy of resonance
states is determined with respect to the $\alpha+\alpha$ threshold energy.%

%TCIMACRO{\TeXButton{B}{\begin{table}[tbp] \centering}}%
%BeginExpansion
\begin{table}[tbp] \centering
%EndExpansion
\caption{Spectrum of resonance states in $^8$Be, calculated with three
different nucleon-nucleon potentials and compared with experimental data.}%
\begin{tabular}
[c]{|c|c|c|c|c|c|}\hline
&  & \multicolumn{2}{|c|}{Theory} & \multicolumn{2}{|c|}{Experiment}\\\hline
$L^{\pi}$ & Potential & $E$, MeV & $\Gamma$, MeV & $E$, MeV & $\Gamma$,
MeV\\\hline\hline
$0^{+}$ & MHNP & 0.091 & 5.183$\cdot$10$^{-6}$ & 0.092 & (5.57$\pm$%
0.25)$\cdot$10$^{-6}$\\
& VP & 0.091 & 6.947$\cdot$10$^{-6}$ &  & \\
& MP & 0.092 & 5.876$\cdot$10$^{-6}$ &  & \\\hline\hline
$2^{+}$ & MHNP & 2.818 & 1.122 & 3.122$\pm$0.010 & 1.513$\pm$0.015\\
& VP & 2.526 & 1.494 &  & \\
& MP & 2.977 & 1.773 &  & \\\hline\hline
$4^{+}$ & MHNP & 10.633 & 1.816 & 11.442$\pm$0.150 & 3.500\\
& VP & 10.852 & 6.732 &  & \\
& MP & 12.710 & 5.281 &  & \\\hline
\end{tabular}
\label{Tab: Resonan8Be}%
%TCIMACRO{\TeXButton{E}{\end{table}}}%
%BeginExpansion
\end{table}%
%EndExpansion

\begin{figure}[hptb]
\begin{center}
\includegraphics[
%natheight=6.373700in,
%natwidth=7.087100in,
%height=11.8859cm,
width=13.2105cm
]{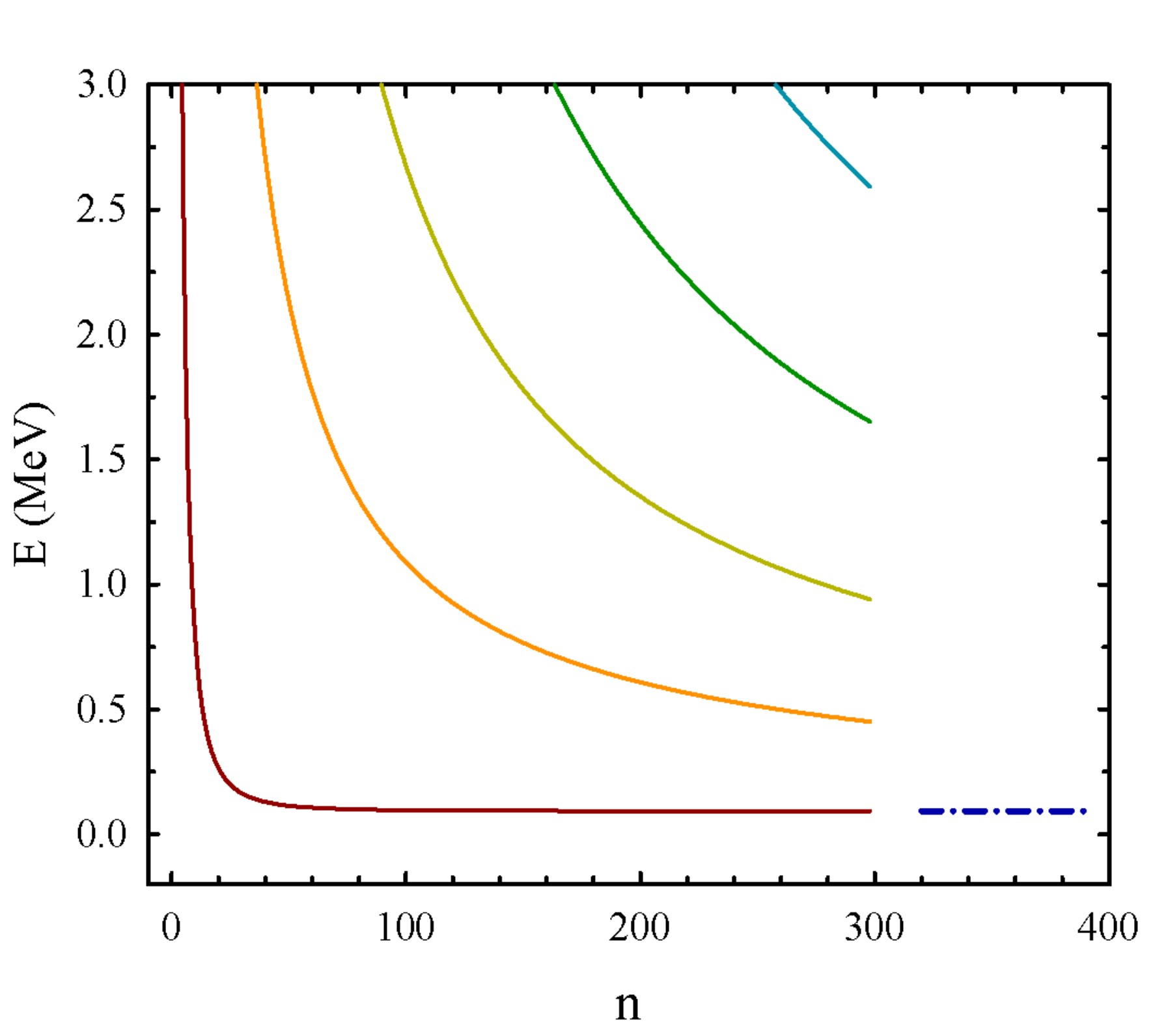}%
%{Spectr8BeJ0MHNP.jpg}%
\caption{Energy of the $0^{+}$ states of $^{8}Be$  as a function of number of
oscillator functions. Dash-doted line indicates position of the $0^{+}$
resonance state. Results are obtained with the MHNP.}%
\label{Fig:Spectr8BeJ0MHNP}%
\end{center}
\end{figure}

\begin{figure}[hptb]
\begin{center}
\includegraphics[
%natheight=6.373700in,
%natwidth=6.947000in,
%height=12.2132cm,
width=13.3071cm
]{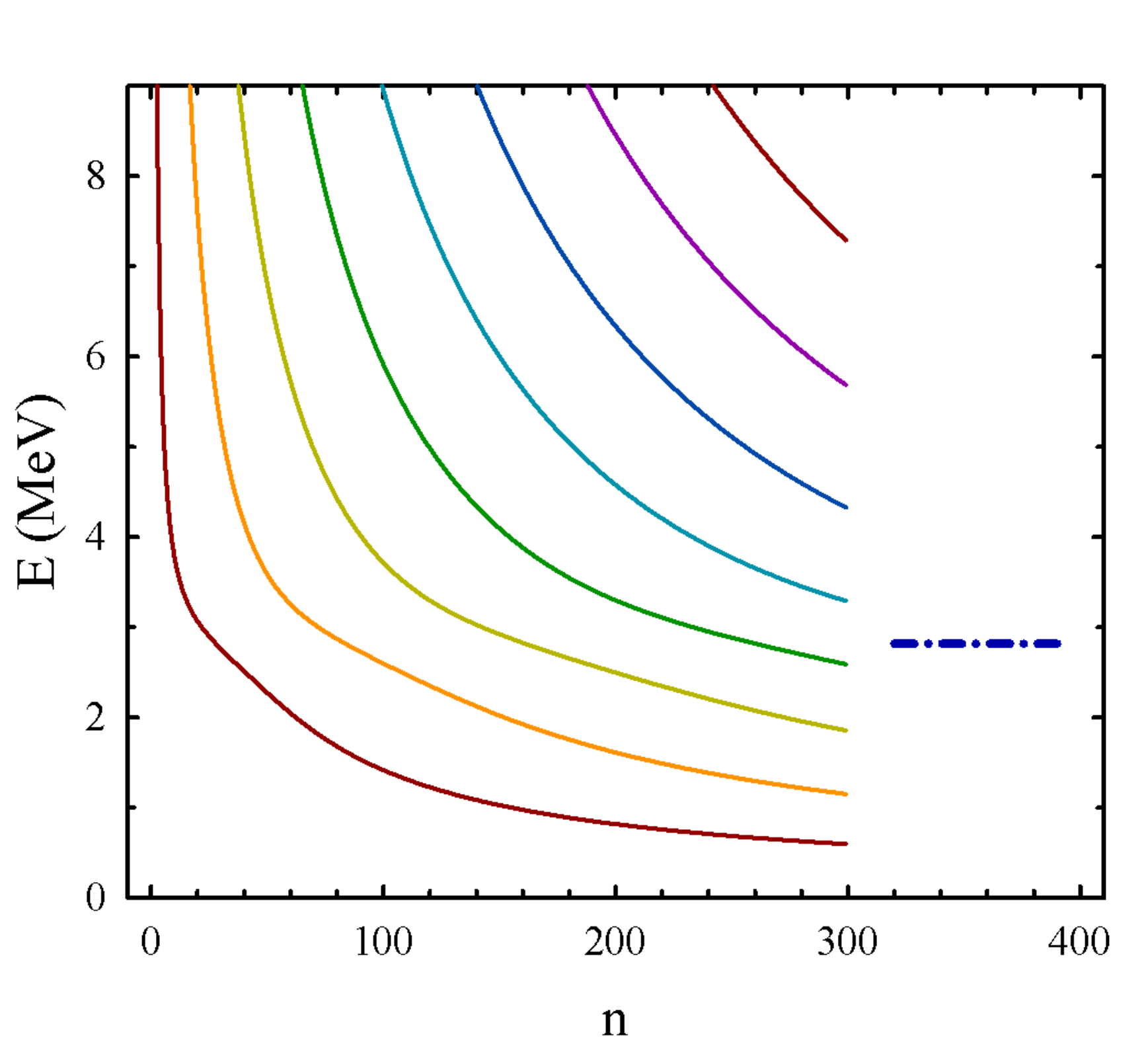}%
%{Spectr8BeJ2MHNP.jpg}%
\caption{Energy of $2^{+}$ states of $^{8}Be$ as a function of number of
oscillator functions. Dash-doted line indicates position of the $2^{+}$
resonance state. Results are obtained with the MHNP.}%
\label{Fig:Spectr8BeJ2MHNP}%
\end{center}
\end{figure}

There is other way to detect resonance states with small width (i.e. narrow
resonance states). This can be done by considering spectrum of two-cluster
Hamiltonian as a function of number of oscillator functions involved in
calculations. Such dependence is shown in Figures \ref{Fig:Spectr8BeJ0MHNP}
and \ref{Fig:Spectr8BeJ2MHNP} where spectra of $0^{+}$ and $2^{+}$ state are
displayed. By dash-doted lines we indicated position of the $0^{+}$ and
$2^{+}$ resonance states, calculated by using the corresponding phase shifts
and equations (\ref{eq:R01}). As we see in\ Figure \ref{Fig:Spectr8BeJ0MHNP},
energy of the lowest $0^{+}$ state has a plateau exactly at the energy of the
$0^{+}$ resonance state. There is no plateau in Figure
\ref{Fig:Spectr8BeJ2MHNP}\ \ for the wide $2^{+}$ resonance state. There are
only small irregularities in behavior of energy of $2^{+}$ states as a
function of number of oscillator functions $N$. Consequently, such type of
figures similar to Figure \ref{Fig:Spectr8BeJ0MHNP} allows one to predict with
very good precision the energy of a very narrow resonance state. This
phenomenon is used in the Stabilization Method (see formulation of the method
in Ref. \cite{1970PhRvA...1.1109H} and some additional illustrations of the
method in \cite{kn:cohstate2E} ) to locate position of resonance states.%

%BeginExpansion
%EndExpansion
%

%BeginExpansion
%EndExpansion

\begin{figure}[hptb]
\begin{center}
\includegraphics[
%natheight=5.513200in,
%natwidth=7.032700in,
%height=10.434cm,
width=13.2918cm
]{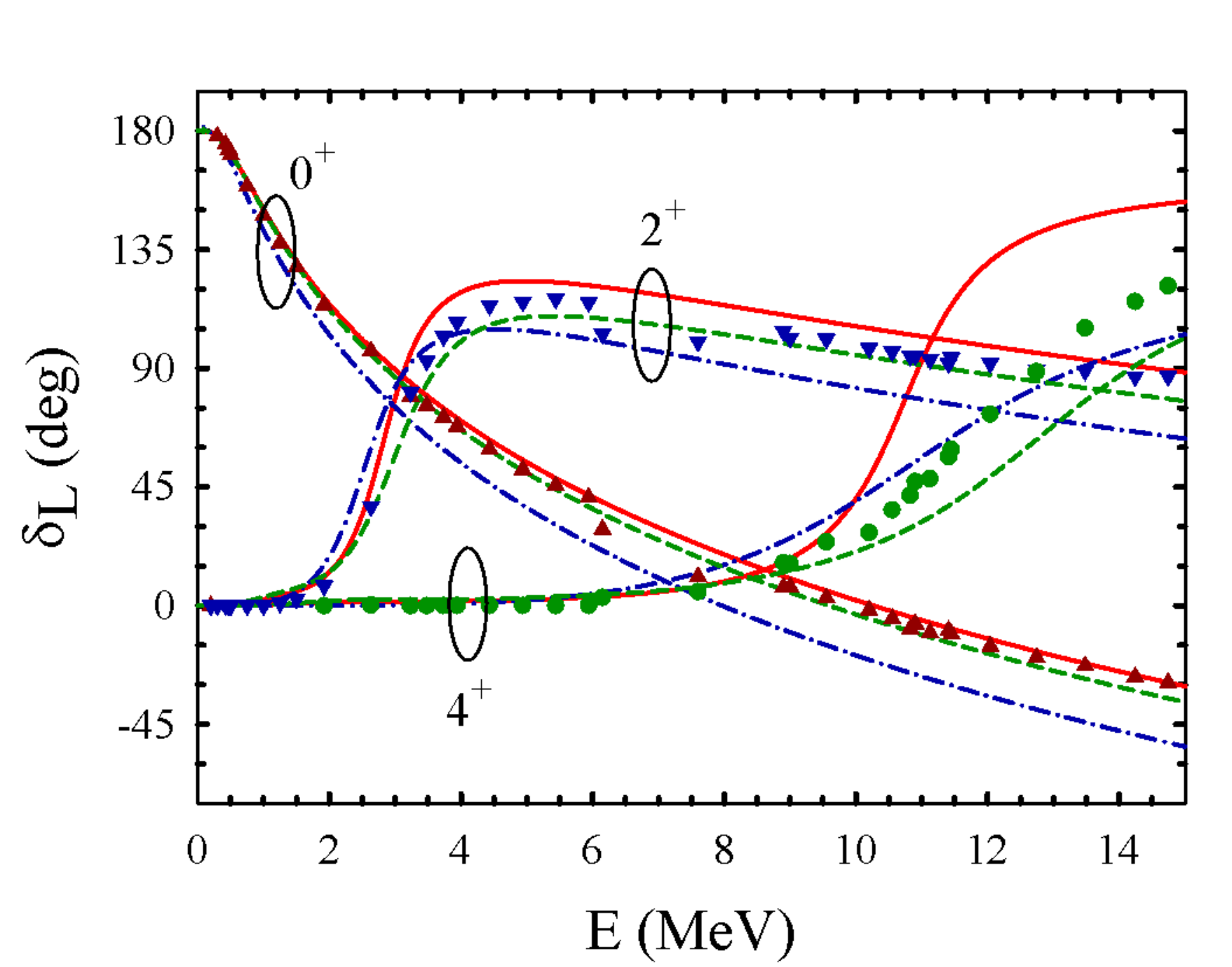}%
%{Phases8BeAllExp.jpg}%
\caption{Phase shifts of the elastic $\alpha+\alpha$ scattering calculated
with three different NN potentials and compared with experimental data.}%
\label{Fig:Phases8BeAllPot}%
\end{center}
\end{figure}

In Figure \ref{Fig:Phases8BeAllPot} we demonstrate dependence of phase shifts
for the alpha-alpha elastic scattering on energy and shape of nucleon-nucleon
potentials. The $S$ phase shifts calculated with the MHNP\ and MP are very
close in the whole range of energies displayed in Fig.
\ref{Fig:Phases8BeAllPot}. However, the $S$ phase shifts calculated with VP
are noticeable different, despite that the VP as the MHNP\ and MP gives the
correct position of the $0^{+}$ resonance state, as it was demonstrated in
Table \ref{Tab: Resonan8Be}. Difference between phases shifts of
$\alpha-\alpha$ scattering generated by three potentials is growing with
increasing of the total orbital momentum $L$. This is also reflected on the
energy and width of the $2^{+}$ and $4^{+}$ resonance states (see Table
\ref{Tab: Resonan8Be}).%

%BeginExpansion
%EndExpansion
Fig. demonstrates that there is fairly good agreement of our results and
experimental data. The experimental data displayed in Fig.
\ref{Fig:Phases8BeAllPot} are taken from Refs. \cite{1956PhRv..104..123H},
\cite{1963PhRv..129.2252T}, \cite{1965PhRv..137..315D},
\cite{1974PhRvC..10.1767C}. One can see that the MHNP\ and MP potentials yield
the $0^{+}$ and $2^{+}$ phase shifts which are very close for experimental
data in the presented range of energy. As for the phase shifts with the total
orbital momentum $4^{+}$, our results are close to the experimental data at
the energy range 0$\leq E\leq$10 MeV, and there is deviation from experimental
data for all potentials for the energy $E>$ 10 MeV.

\begin{figure}[hptb]
\begin{center}
\includegraphics[
%natheight=6.373700in,
%natwidth=7.073300in,
%height=11.8859cm,
width=13.1841cm
]{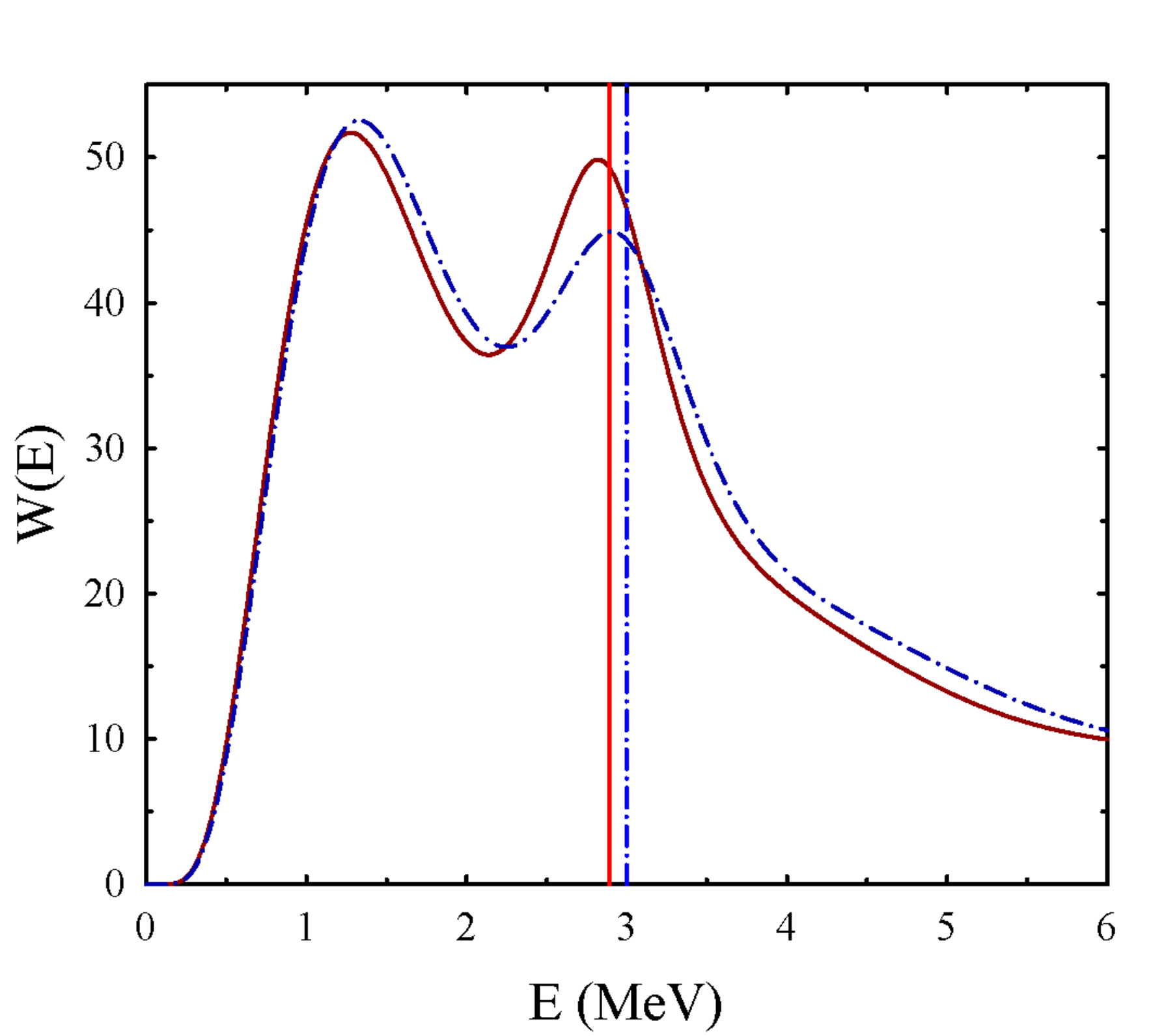}%
%{Weights8BeL2.jpg}%
\caption{The weights of internal part of scattering wave function as a
function of energy constructed for the $2^{+}$ state with the MHNP\ (solid
line) and MP (dot-dashed line). Vertical lines indicate position of the
$2^{+}$ resonance states.}%
\label{Fig:Weights8BeL2}%
\end{center}
\end{figure}

The weights $W\left(  E\right)  $ of the internal part of scattering wave
function introduced in Eq. (\ref{eq:M21}) also reflects existence of resonance
state. We demonstrate it for the $2^{+}$ states. The weights displayed in Fig.
\ref{Fig:Weights8BeL2} are calculated with the MHNP\ (solid line) and with the
MP\ (dot-dashed line). The solid vertical and dot-dashed lines indicates the
position of the $2^{+}$ resonance states for the MHNP and MP, respectively.
There are two picks in Fig. . The first peak does not connected with resonance
state, while the second peak is formed by resonance state. The center of the
peak are very close to the energy of the $2^{+}$ resonance state. This figure
demonstrates that the weights $W\left(  E\right)  $ as function of energy can
confirm existence of rather wide resonance state and indicate of its
position.
%BeginExpansion
%EndExpansion

\subsection{Resonance wave functions}

\begin{figure}[hptb]
\begin{center}
\includegraphics[
%natheight=8.553000in,
%natwidth=6.406500in,
%height=17.6674cm,
width=12.90cm
]{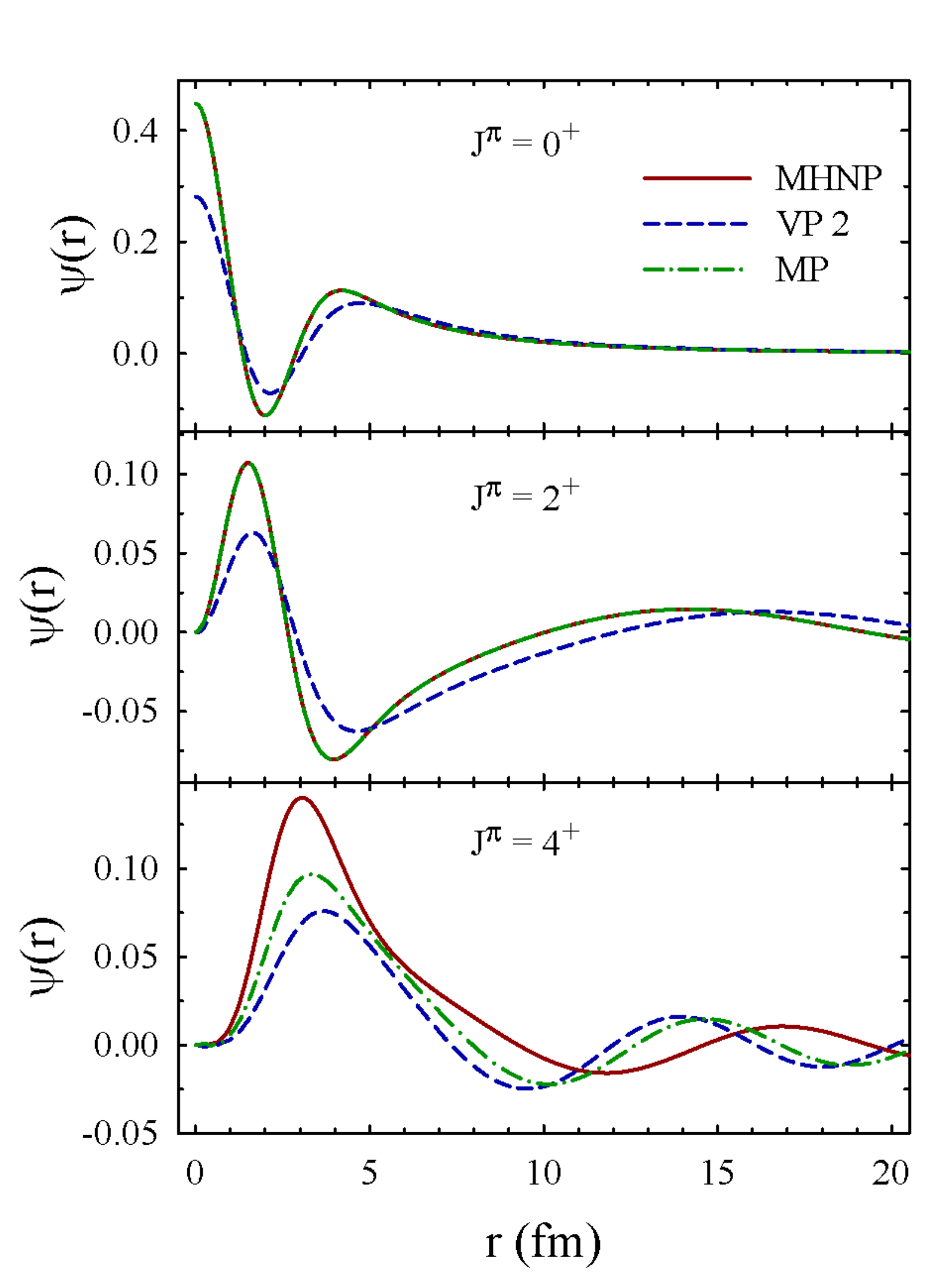}%
%{ResonWaveFuns8BeN.jpg}%
\caption{Wave functions of the $0^{+}$, $2^{+}$ and $4^{+}$ resonance states
as a function of distance between alpha particles, calculated with the MHNP,
VP and MP potentials.}%
\label{Fig:WaveFuns8BeRS}%
\end{center}
\end{figure}

In this section we consider wave functions of resonance states in $^{8}$Be. We
present these functions in oscillator, coordinate and momentum
representations. We start with coordinate representations. The wave functions
$\psi_{EL}\left(  r\right)  $ of the $0^{+}$, $2^{+}$ and $4^{+}$ resonance
states are displayed in Figure \ref{Fig:WaveFuns8BeRS}. \ It is interesting to
note that maximum of the inter-cluster wave function of the $0^{+}$ resonance
state is at $\ r=0$. Thus the Pauli principle allows two alpha particles to be
at the same point of the coordinate space. Due to the centrifugal barrier,
wave functions of the $\ 2^{+}$ and $4^{+}$ resonance states \ equal zero at
$r=0$. Other interesting feature of all resonance states is that they have
large amplitude of wave function in the internal region ($0\leq r<7$ fm) and
small amplitude of oscillations in asymptotic region ($r>7$ fm). Presented
results allows us to investigate dependence of resonance wave functions on
shape of nucleon-nucleon potential. As we can see in Figure
\ref{Fig:WaveFuns8BeRS}, the MHNP and MP give almost identical wave functions
for the $0^{+}$ and $2^{+}$ resonance states, but slightly different wave
functions for the very broad $4^{+}$ resonance state. Besides, the amplitude
of resonance wave functions, calculated with the MHNP and MP, is larger than
amplitude of the wave functions, obtained with the Volkov potential.%

%BeginExpansion
%EndExpansion

\begin{figure}[hptb]
\begin{center}
\includegraphics[
%natheight=8.426700in,
%natwidth=6.260400in,
%height=17.8344cm,
width=13.2676cm
]{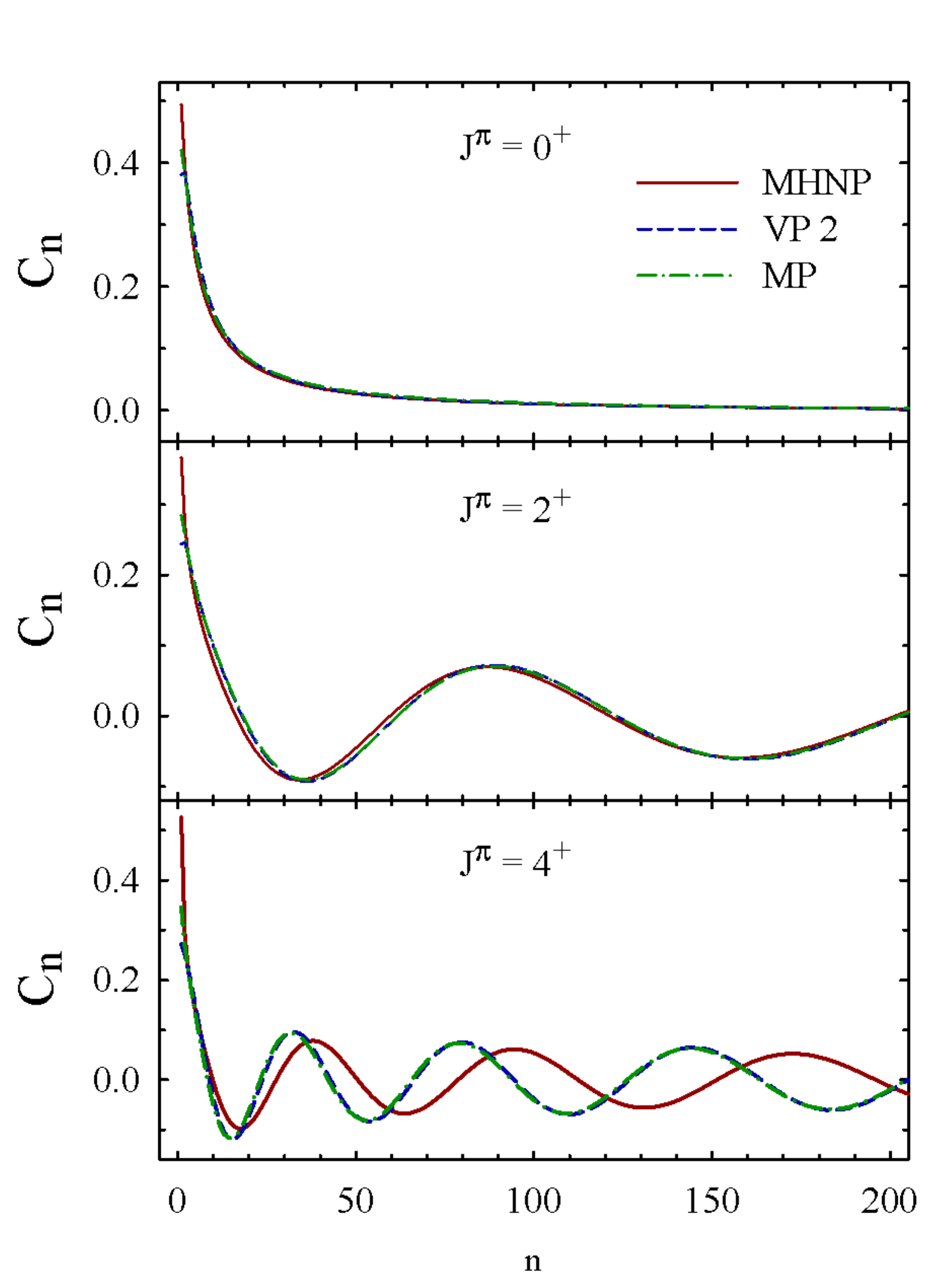}%
%{ResonExpCoeff8Be.jpg}%
\caption{Wave functions of $0^{+}$, $2^{+}$ and $4^{+}$ resonance states in
$^{8}$Be in oscillator representation.}%
\label{Fig:ExpCoeff8BeRS}%
\end{center}
\end{figure}

In Figure \ref{Fig:ExpCoeff8BeRS} we present wave function of the $0^{+}$,
$2^{+}$ and $4^{+}$ resonance states in the oscillator representation. Wave
functions of the narrow $0^{+}$resonance states look like wave function of
bound states. Our results confirm conclusion made in Ref.
\cite{2015NuPhA.941..121L}, that wave function of narrow resonance states in
coordinate or oscillator representations are similar to wave functions of
bound states. As we can see, that oscillator functions with \ 0$\leq$n$<$20
dominate in wave functions of all observed resonance states, since\ expansion
coefficients $C_{n}$, associated with these functions, have largest weight in
resonance wave functions. Resonance wave functions in oscillator
representation are similar to corresponding wave functions in coordinate
representation: they have large amplitude in the internal region.

Figure \ref{Fig:ExpCoeff8BeRS} demonstrates that the wave functions of the
$0^{+}$ and $2^{+}$ resonance states almost independent on shape of
nucleon-nucleon potential. The main difference between wave functions of
different nucleon-nucleon potentials is observed for the expansion
coefficients with very small values of $n$. As for the wave functions of the
$4^{+}$ resonance state, they are very close to the VP and MP potentials,
while the wave functions of the resonance states obtained with the MHNP is
quite different.

Analysis of results, presented in Figures \ref{Fig:WaveFuns8BeRS} and
\ref{Fig:ExpCoeff8BeRS}, shows that the larger is the energy of resonance
state, the more oscillations has resonance wave function in coordinate and
oscillator representations within the displayed range of variables $r$ and
$n$, respectively.%

%BeginExpansion
%EndExpansion

\begin{figure}[hptb]
\begin{center}
\includegraphics[
%natheight=8.633400in,
%natwidth=6.154000in,
%height=18.489cm,
width=12.25cm
]{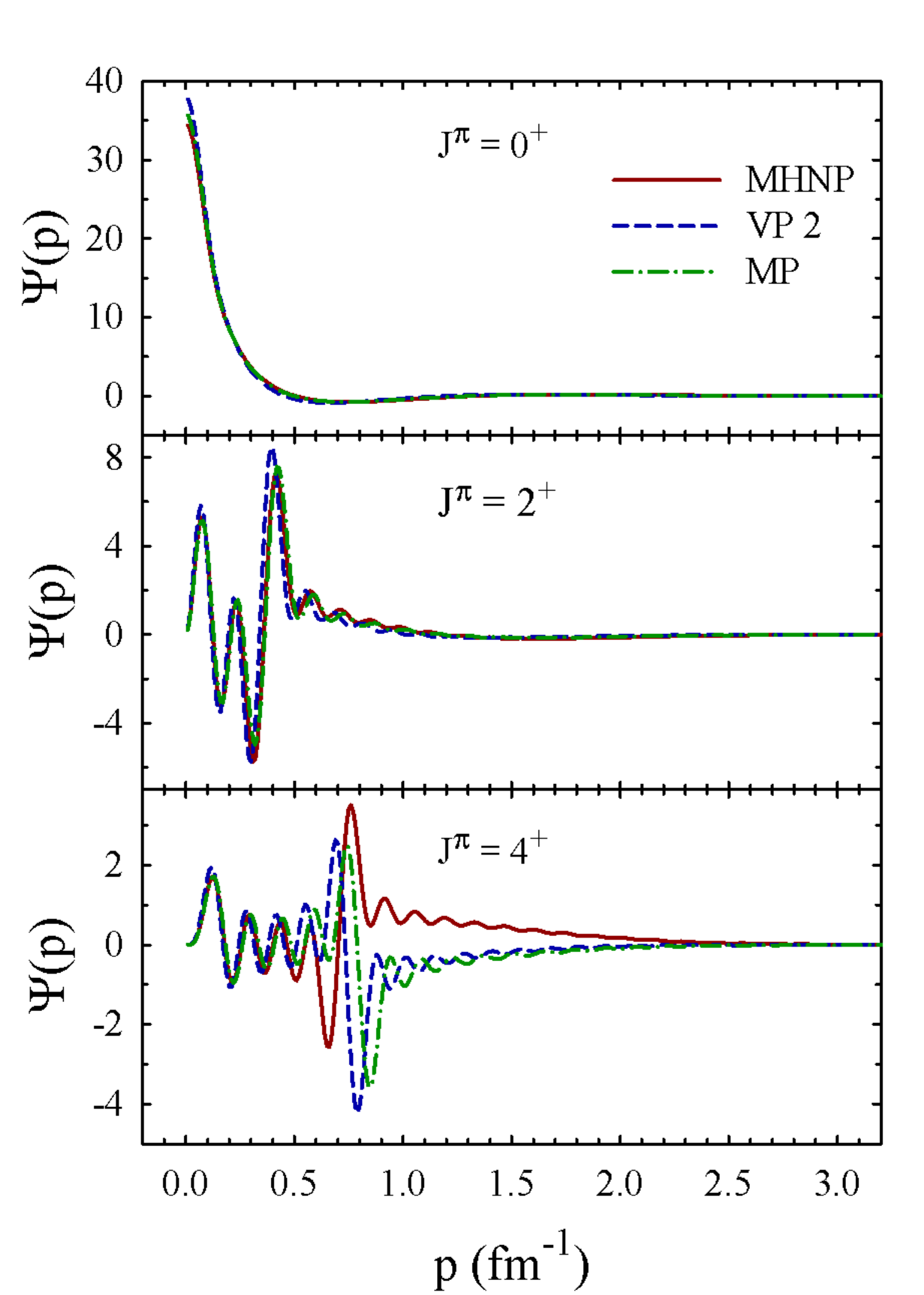}%
%{ResonWaveFuns8BeMS.jpg}%
\caption{Wave functions of the $0^{+}$, $2^{+}$ and $4^{+}$ resonance states
in $^{8}$Be in the momentum space. Results are presented for three NN
potentials.}%
\label{Fig:WaveFunsRS8BeMS}%
\end{center}
\end{figure}

Wave functions of resonance states in $^{8}$Be are also presented in momentum
space (Figure \ref{Fig:WaveFunsRS8BeMS}). Figure demonstrates that wave
functions of $0^{+}$ and $2^{+}$ resonance state are almost independent on the
shape of the nucleon-nucleon potential. These wave functions are located in
very restricted region of the momentum $p$. Wave functions of the $4^{+}$
resonance states, obtained with different nucleon-nucleon potentials, are
quite different and they spread in more wider region of $p$. However, they are
similar in the range of small values of momentum: $0\leq p\leq0.5$ fm$^{-1}$.

%BeginExpansion
%EndExpansion

To demonstrate that our method is consistent with other alternative methods,
we compare our results with results of the Complex Scaling Method (CSM)
(recent achievements of the method and its main ideas can be found in Ref.
\cite{2014PrPNP..79....1M}) which allows one to determine pole of the
$S$-matrix and consequently energy and width of resonance states directly from
calculations of eigenvalues of the nonhermitian Hamiltonian. Both calculations
are performed with the MHNP and with the same input parameters. Results of
these calculations are gathered in Table \ref{Tab:RGMvsCSM}. It is seen that
results of our calculations are quite comparable with results of the CSM. The
best agreement between two methods is achieved for the $2^{+}$ and $4^{+}$
resonance states, meanwhile energy and width of the $0^{+}$ resonance state
are surprisingly different within these methods.%

%TCIMACRO{\TeXButton{B}{\begin{table}[htbp] \centering}}%
%BeginExpansion
\begin{table}[htbp] \centering
%EndExpansion
\caption{Parameters of resonance states calculated within the Resonating
Group Method and the Complex Scaling Method.}%
\begin{tabular}
[c]{|c|c|c|c|c|}\hline
& \multicolumn{2}{|c|}{Our method} & \multicolumn{2}{|c|}{CSM}\\\hline
$J^{\pi}$ & $E$, MeV & $\Gamma$, MeV & $E$, MeV & $\Gamma$, MeV\\\hline
$0^{+}$ & 0.091 & 5.183$\cdot$10$^{-6}$ & 0.150 & 35.68$\cdot$10$^{-5}%
$\\\hline
$2^{+}$ & 2.820 & 1.196 & 2.893 & 1.135\\\hline
$4^{+}$ & 10.730 & 1.925 & 10.824 & 1.916\\\hline
\end{tabular}
\label{Tab:RGMvsCSM}%
%TCIMACRO{\TeXButton{E}{\end{table}}}%
%BeginExpansion
\end{table}%
%EndExpansion

\begin{figure}[hptb]
\begin{center}
\includegraphics[
%natheight=6.406500in,
%natwidth=7.486700in,
%height=11.2994cm,
width=13.1907cm
]{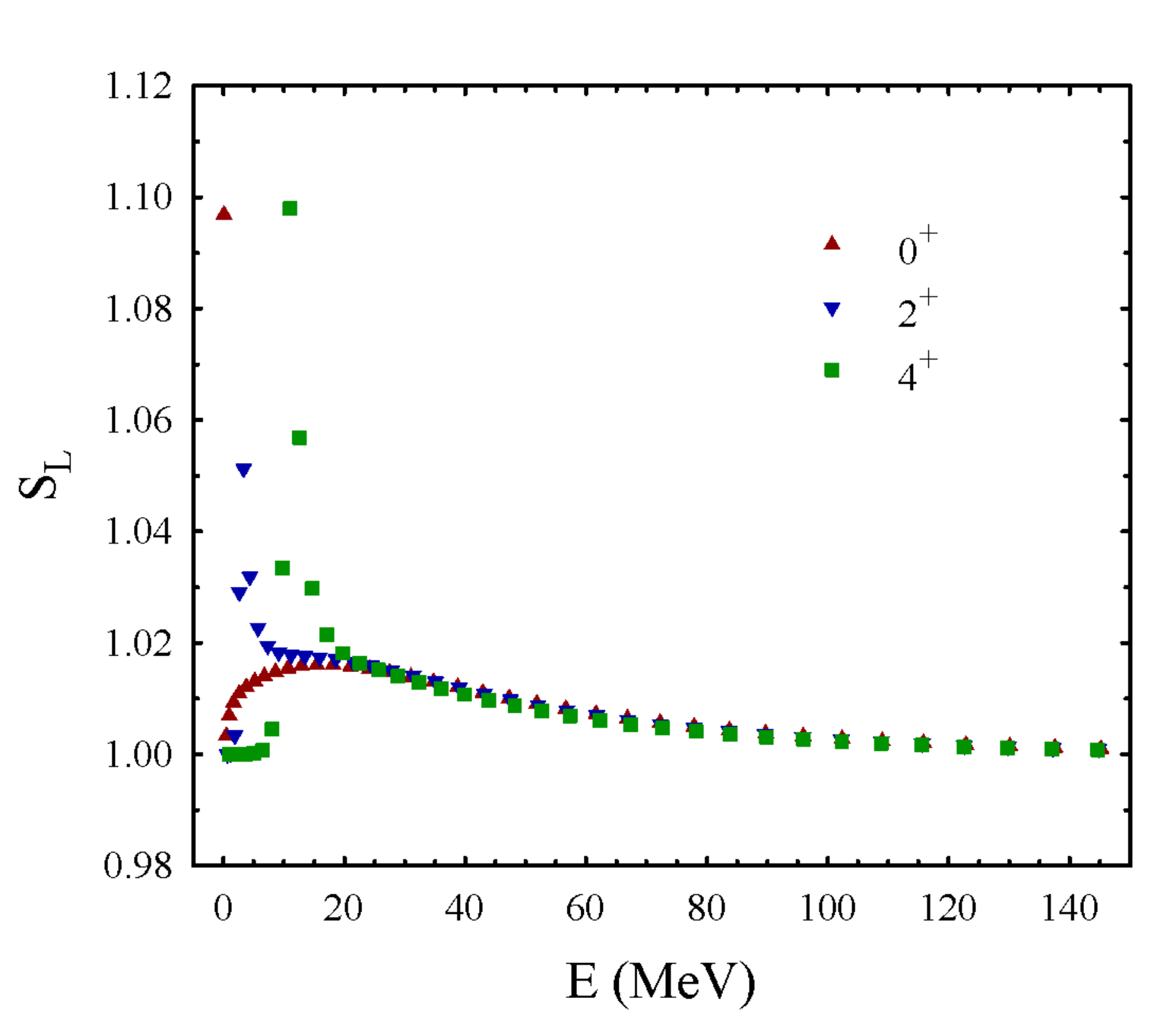}%
%{SJFactor8Be.jpg}%
\caption{Effects of the Pauli principle in the wave functions of Hamiltonian
of $^{8}$Be.}%
\label{Fig:SJFactor8Be}%
\end{center}
\end{figure}

It is well-known that the Pauli principle plays an important role in many-body
fermion systems and particular in many-cluster systems. \ The effects have
been discussed from a different point of view. We present other way of
discussing role of the Pauli principle on wave functions of continuous
spectrum states in $^{8}$Be. We calculate and analyze the quantity $S_{L}$
introduced in Eq. (\ref{eq:C19A}). This quantity equals to unity if the Pauli
is disregarded or if its effects are negligibly small. Thus, deviation of
$S_{L}$ from unity shows how strong is the effect of the Pauli principle. In
Fig. \ref{Fig:SJFactor8Be} we display $S_{L}$ \ as a function of energy for
the $0^{+}$, $2^{+}$ and $4^{+}$ states.%

%BeginExpansion
%EndExpansion
We can see that the largest effect of the Pauli principle is observed for the
eigenfunctions whose energy is very close to the energies of detected
resonance states. This results reflects the main properties of resonance
states, which are the most compact configurations among other states of
two-cluster continuum. The smaller is the width of a resonance state, the more
compact is its configuration. Fig. \ref{Fig:SJFactor8Be} demonstrate that the
low-energy states (0$\leq E\leq$50 MeV) are strongly affected by the Pauli
principle than the states with large energy. With increasing of the energy the
effects are gradually (steadily) diminished. It is worth noticing that the
huge centrifugal barrier in the 4$^{+}$ states strongly diminished effects of
the Pauli principle for the low-energy states (0$\leq E\leq$5 MeV) where
$S_{J}\approx$1.The centrifugal barrier in 2$^{+}$ states \ is approximately 3
times smaller than in 4$^{+}$ states and thus the Pauli principle is stronger
for 2$^{+}$ states in this range of energy. It is interesting to note that the
results shown in Fig. \ref{Fig:SJFactor8Be} are in agreement with conclusions
made in Ref. \cite{2015NuPhA.941..121L} where two-cluster systems including
$^{8}$Be have been studied in the Fock-Bargmann representations. This
representation is a bridge between quantum-mechanical treatment of nuclear
system and classical. It was in particular shown that the classical regime is
valid in continuous spectrum of $^{8}$Be when energy of such states is larger
than 60 MeV. Fig. \ref{Fig:SJFactor8Be} also shows that effects the Pauli
principle in this region is small.

\begin{figure}[hptb]
\begin{center}
\includegraphics[
%natheight=6.406500in,
%natwidth=7.360400in,
%height=11.6245cm,
width=13.3445cm
]{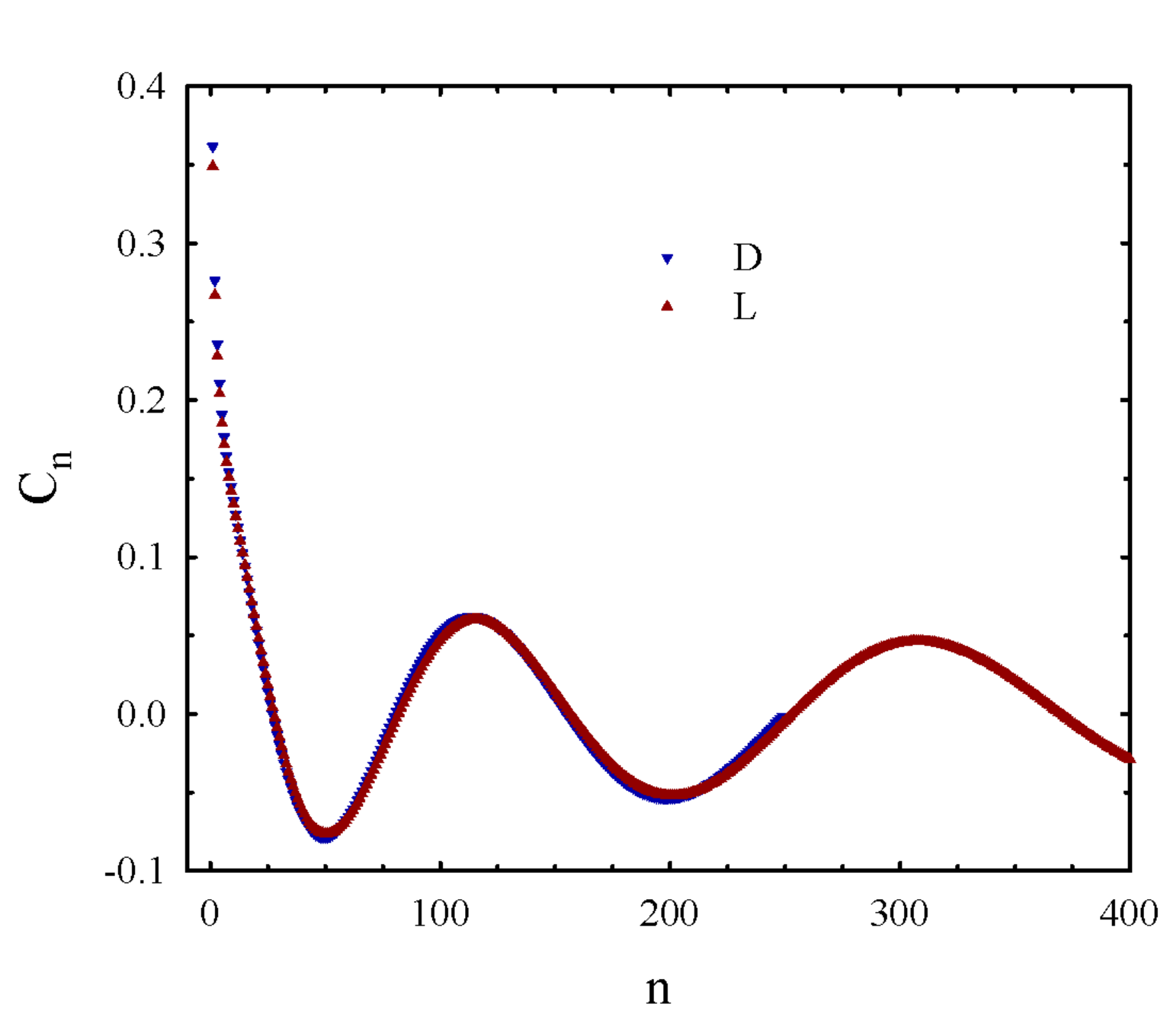}%
%{WaveFun8BeL2DvsL.jpg}%
\caption{Wave functions of the $2^{+}$ state obtained with diagonalization of
Hamiltonian ($D$) and by solving system of linear algebraic equations ($L$).}%
\label{Fig:WaveFun8BeL2LvsD}%
\end{center}
\end{figure}

To demonstrate relation between wave function obtained with diagonalization of
Hamiltonian with a certain number of oscillator functions and wave function
calculated by solving a set of linear equations (\ref{eq:M13}) with necessary
boundary conditions, we used the following procedure. First, we diagonalize
250$\times$250 matrix of Hamiltonian for the 2+ state and select an eigenstate
with the energy close to the energy of $2^{+}$ resonance states. Thus we
selected the four eigenfunction with the energy $E_{4}$=2.929 MeV. This
function we mark as function D, as it was obtained with the diagonalization.
Second, we solved the system of linear algebraic equations with the energy
very close to the energy of the four eigenstates. We employ 400 oscillator
functions to show behavior of expansion coefficients in a large range of index
$n$. The wave function calculated in such way is marked by liter L. As they
normalized in different ways, we renormalize wave function L to make
comparison self-consistent. We divided wave function L\ $\left\{
C_{nL}^{\left(  E\right)  }\right\}  $ on the square root of the sum $S_{250}$%
\[
S_{250}=\sum_{n=0}^{249}\left\vert C_{nL}^{\left(  E\right)  }\right\vert ^{2}%
\]
As the results part of wave function L represented by 250 oscillator functions
is normalized in the same way as function D. Renormalized wave function L and
wave function D are demonstrated in Fig. \ref{Fig:WaveFun8BeL2LvsD}. There is
small difference for the expansion coefficients with small values of $n$,
which owned to small difference of energy of $D$ and $L$ states. In a large
range of $n$ they are very close to each other. Wave function $L$ can be
easily extended to very large values $n$ ($n$ >
400) since we know an asymptotic behavior of the function.

%BeginExpansion
%EndExpansion

\begin{figure}[hptb]
\begin{center}
\includegraphics[
%natheight=6.406500in,
%natwidth=7.406300in,
%height=11.4598cm,
width=13.239cm
]{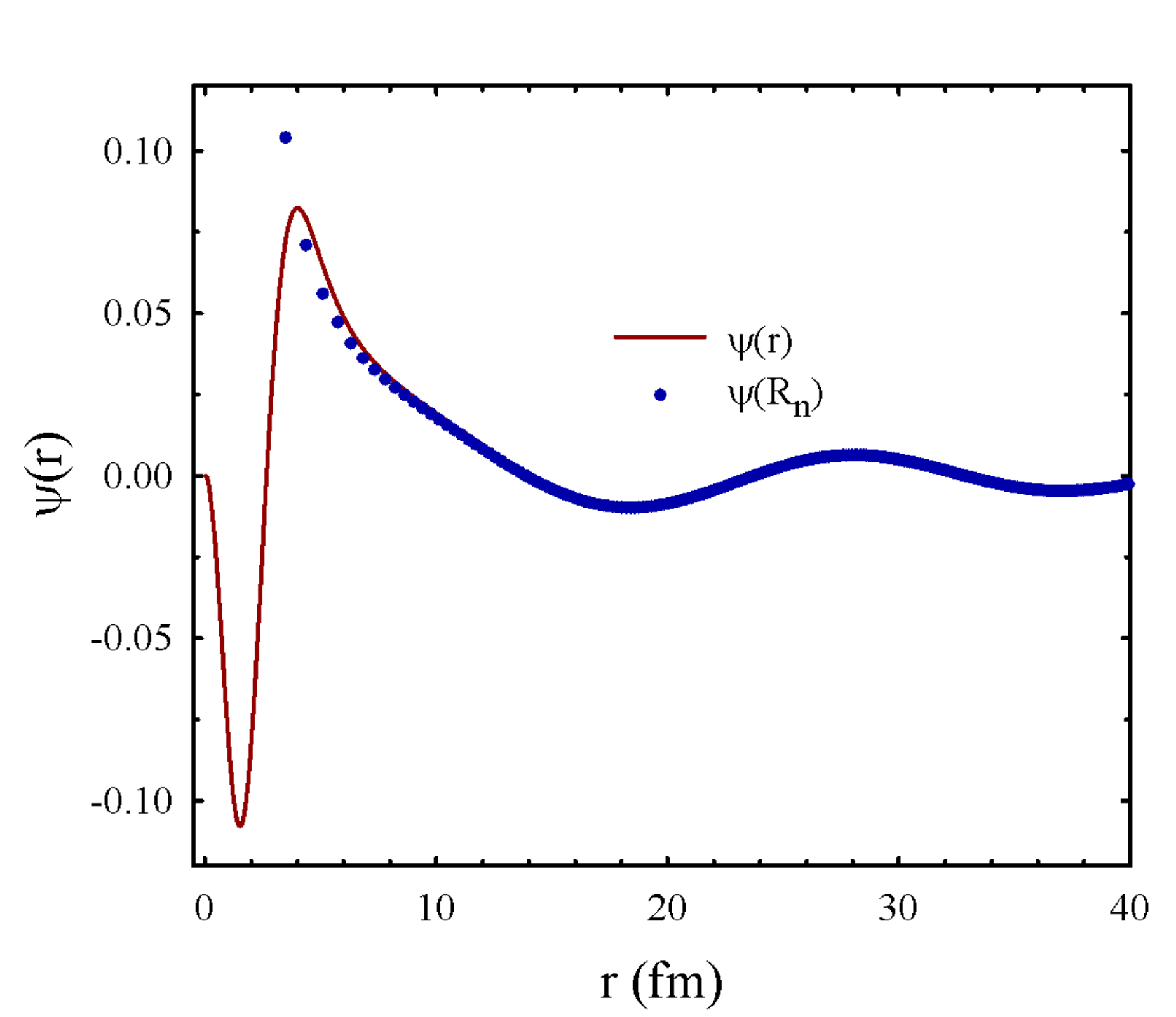}%
%{WaveFun8BeL2DvsC.jpg}%
\caption{Wave function $\psi\left(  r\right)  $ of the 2+ state with the
energy $E$=2.92 MeV as function of inter-cluster distance (solid line) and
expansion coefficients ($\psi\left(  R_n\right)  $) of the function as a
function of discrete distance $R_n$.}%
\label{Fig:WaveFun8BeL2CvsD}%
\end{center}
\end{figure}

We select the wave function of $2^{+}$ state which is displayed in Fig.
\ref{Fig:WaveFun8BeL2LvsD} and obtained from diagonalization procedure, to
show the relation between coordinate wave function and expansion coefficients
discussed above and presented in Eq. (\ref{eq:C23}). Fig.
\ref{Fig:WaveFun8BeL2CvsD} visualize such relation. We can see that all
expansion coefficients starting from $n=4$ are very close to the wave function
in coordinate space. This figure confirms once more that there is simple
connection between coordinate and oscillator representations. This connection
is valid not only for large values of index $n$ as it \ was deduced
originally, but also a fairly small values.%

%BeginExpansion
%EndExpansion

\subsubsection{Average distances between clusters}

In Table \ref{Tabl:AveragDist} we display energy of obtained by diagonalizing
Hamiltonian, mass root-mean-square (rms) radii \ $R_{m}$ of $^{8}$Be in these
states, and average distance $A_{c}$ between clusters. These results are
obtained with MHNP. It is necessary to point out that the most compact
configurations of $^{8}$Be are revealed in those eigenstates of Hamiltonian
whose energies are very close to the energy of resonance states. They are the
first state for $L^{\pi}=0^{+}$ and the fourth state for $L^{\pi}=2^{+}$.
These two states have the smallest mass rms radius and smallest average
distance between alpha particles. As we can see, the average distances are
approximately two times larger than corresponding mass rms radii. It is also
worth noticing that the low-energy states (except the 0$^{+}$ resonance state)
have largest distance between alpha-particles. Thus \ the low-energy states in
a cluster model are very dispersed (stretched) states. This is stipulated by
the shape of their wave functions which dominantly represented by oscillator
functions with large values of the index $n$.%

%TCIMACRO{\TeXButton{B}{\begin{table}[tbp] \centering}}%
%BeginExpansion
\begin{table}[tbp] \centering
%EndExpansion
\caption{The mass root-mean-square radii and average distances between alpha particles 
for some pseudo-bound $0^{+}$, $2^{+}$ and $4^{+}$ states.}%
\begin{tabular}
[c]{|c|c|c|c|c|c|c|c|c|}\hline
\multicolumn{3}{|c|}{$L^{\pi}=0^{+}$} & \multicolumn{3}{|c|}{$L^{\pi}=2^{+}$}
& \multicolumn{3}{|c|}{$L^{\pi}=4^{+}$}\\\hline
$E$, MeV & $R_{m}$, fm & $A_{c}$, fm & $E$, MeV & $R_{m}$, fm & $A_{c}$, fm &
$E$, MeV & $R_{m}$, fm & $A_{c}$, fm\\\hline
0.091 & 3.11 & 6.35 & 0.610 & 11.31 & 22.62 & 6.44 & 9.67 & 19.35\\\hline
0.462 & 10.74 & 21.48 & 1.168 & 10.14 & 20.28 & 8.07 & 9.44 & 18.88\\\hline
0.953 & 10.01 & 20.02 & 1.876 & 9.49 & 18.99 & 9.71 & 8.62 & 17.25\\\hline
1.664 & 9.89 & 19.78 & 2.624 & 8.25 & 16.51 & 11.03 & 7.59 & 15.24\\\hline
2.604 & 9.84 & 19.69 & 3.329 & 8.32 & 16.67 & 12.60 & 8.93 & 17.87\\\hline
\end{tabular}
\label{Tabl:AveragDist}%
%TCIMACRO{\TeXButton{E}{\end{table}}}%
%BeginExpansion
\end{table}%
%EndExpansion

\begin{figure}[hptb]
\begin{center}
\includegraphics[
%natheight=8.553000in,
%natwidth=7.260100in,
%height=15.4928cm,
width=13.1622cm
]{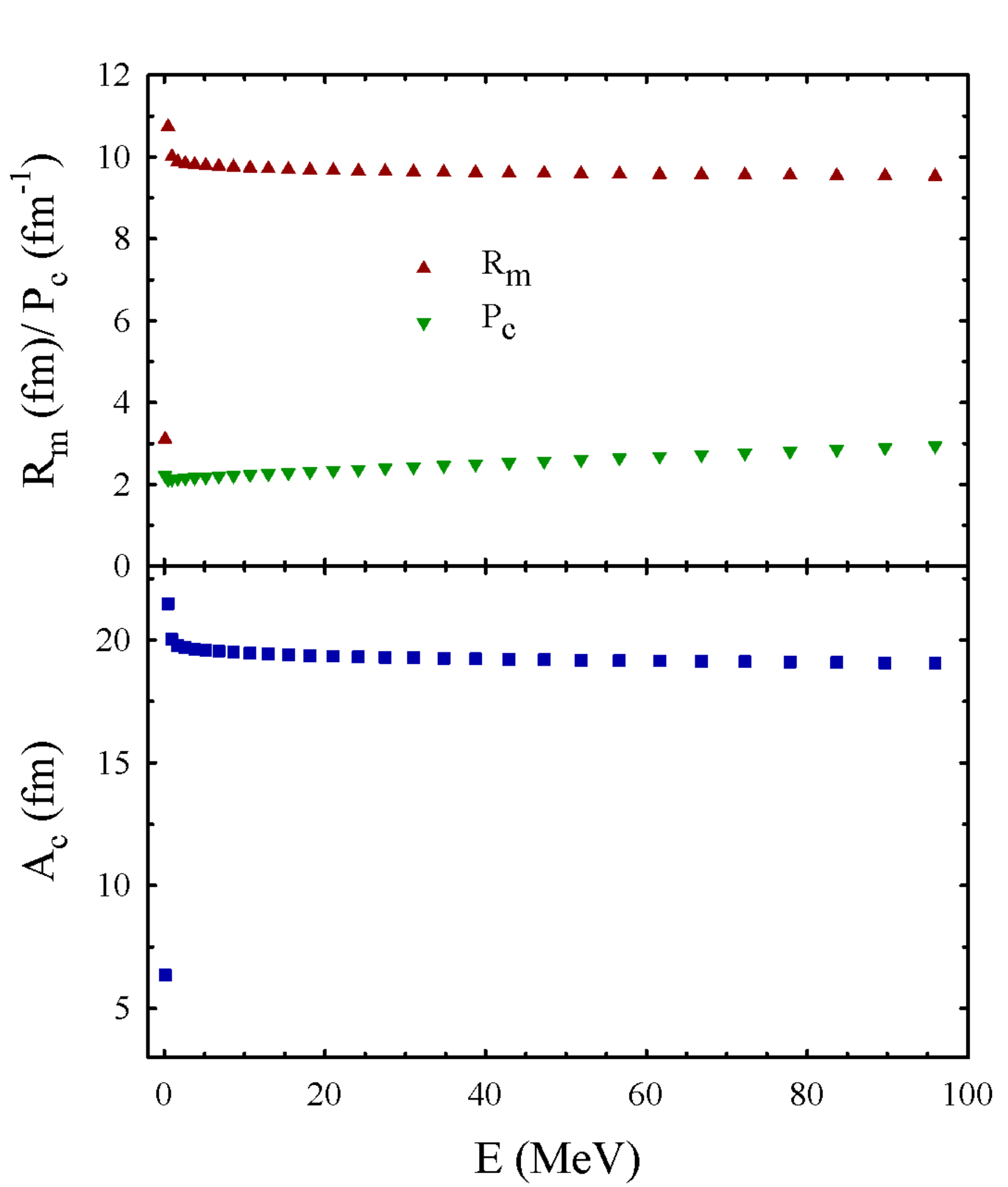}%
%{AveragDist8BeL0.jpg}%
\caption{The mass rms radius $R_{m}$, the average distance between alpha
clusters $A_{c}$ and the average momentum of their relative motion $P_{c}$ as
a function of energy obtained for the $0^{+}$ states with the MHNP.}%
\label{Fig:AveragDist8BeL0}%
\end{center}
\end{figure}

\begin{figure}[hptb]
\begin{center}
\includegraphics[
%natheight=8.553000in,
%natwidth=7.260100in,
%height=15.4928cm,
width=13.1622cm
]{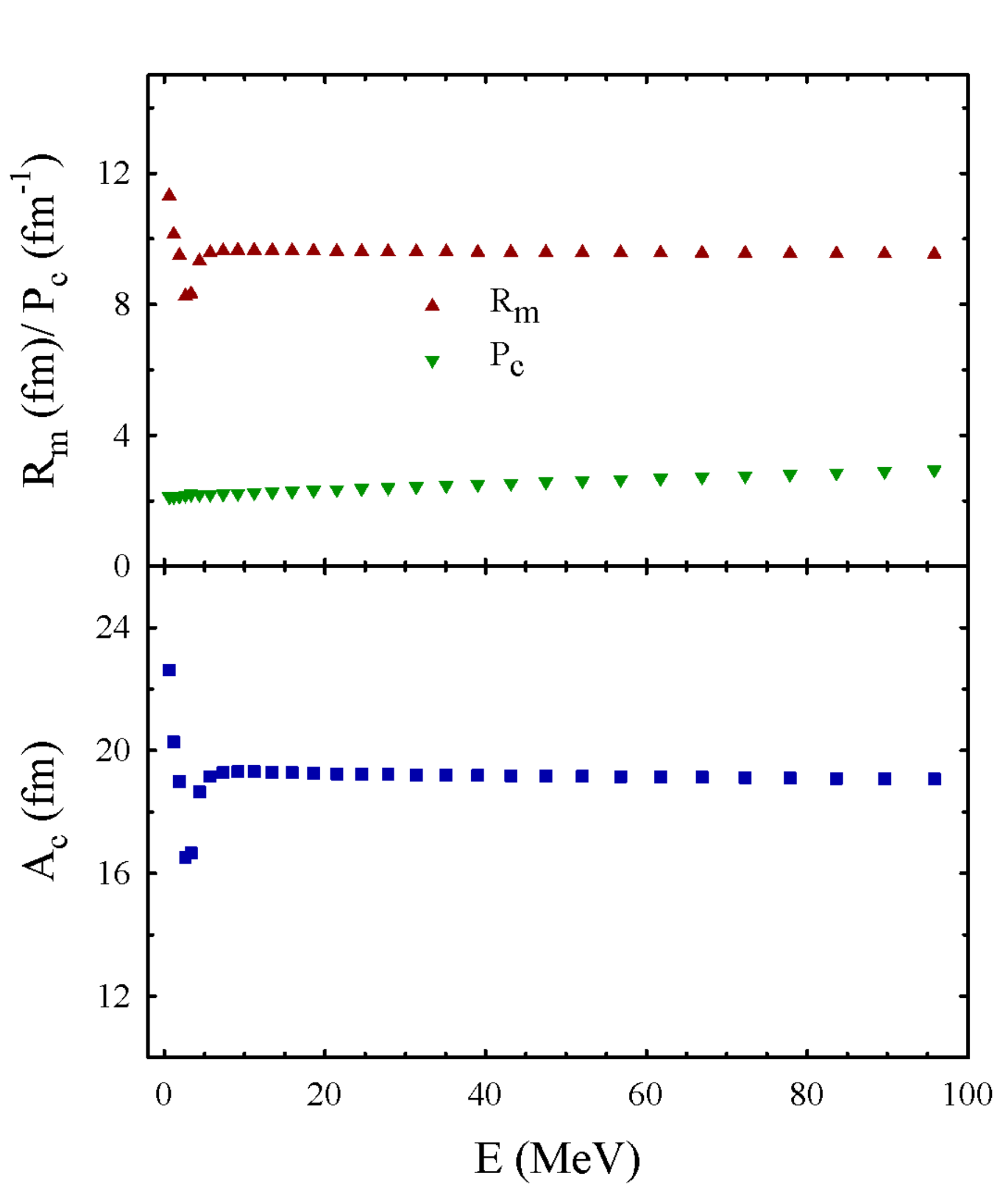}%
%{AveragDist8BeL2.jpg}%
\caption{The same as in Fig. \ref{Fig:AveragDist8BeL0} but for the states
with$\ J^{\pi}$=$2^{+}$}%
\label{Fig:AveragDist8BeL2}%
\end{center}
\end{figure}

In Table \ref{Tabl:AveragDist} we display five $0^{+}$ and $2^{+}$ states with
the lowest energies, while for the $4^{+}$ states we selected those
eigenstates which are close to the energy of the $4^{+}$ resonance state.
Results presented in Table \ref{Tabl:AveragDist} show effects of the Pauli
principle on restricted set of eigenstates. The full picture of the effects
for a large range of states are displayed in Figs. \ref{Fig:AveragDist8BeL0}
and \ref{Fig:AveragDist8BeL2}. In Fig. \ref{Fig:AveragDist8BeL0} we show for
the $0^{+}$ states the mass rms radii $R_{m}$, the average distance between
alpha clusters $A_{c}$ and the average momentum $P_{c}$ as function of energy.
The same information is displayed in Fig. \ref{Fig:AveragDist8BeL2} for the
$0^{+}$ states.%

%BeginExpansion
%EndExpansion
%

%BeginExpansion
%EndExpansion

The average distances between an interacting clusters has been discussed in
literature where cluster models are applied to study different nuclear
hypernuclear system. For example, in Ref. \cite{2018PhRvC..97b4330K} the
hypernucleus $_{\Lambda}^{9}$Be has been studied within a three-cluster model
with the partition $\alpha+\alpha+\Lambda$. The nucleus $^{8}$Be comprised by
two alpha particles is a major ingredient of the model. An approximate formula
was used in Ref. \cite{2018PhRvC..97b4330K} to extract the average
$\alpha-\alpha$ distance. Recall that we use a rigorous way for obtaining this
quantity. The average distance between alpha particles was evaluated to be
5.99\ fm in the ground $0^{+}$ state which is close to our evaluation $A_{c}=$
6.35 fm.

\section{Conclusions \label{Sec:Conclusions}}

We have applied a two-cluster microscopic model for investigation of
alpha-alpha scattering and resonance structure of $^{8}$Be. The model is the
resonating group method with a matrix form of dynamic equations. Transition
from coordinate form to the matrix form was implemented with the help of
oscillator functions. Three popular semi-realistic nucleon-nucleon potentials
were involved in calculations to determine both the internal energy of each
cluster and interaction between them. They also determine dynamics of
alpha-alpha scattering and resonance structure of $^{8}$Be. These potentials
were used to study dependence of \ spectrum, phase shifts and wave functions
of different states on the shape of nucleon-nucleon potential. Small tuning of
parameters of nucleon-nucleon potentials allowed us to reproduce with a good
accuracy the energy and width of the $0^{+}$ resonance state. The same
parameters of potentials were used to calculate $0^{+}$, $2^{+}$ and $4^{+}$
phase shifts \ of elastic alpha-alpha scattering and the position of $2^{+}$
and $4^{+}$. Our results are in a fairly good agreement with available
experimental data and with results of other microscopical models.

It was shown that the Pauli principle has largest impact on the wave functions
of resonance states, since resonance states are the most compact two-cluster
configurations among other states of continuous spectrum. This was confirmed
by calculations of the mass root-mean-square radius and average distance
between alpha-particles. These calculations were performed within large but
finite set of oscillator functions. It was also demonstrated that the effects
of the Pauli principle are steadily decreasing with increasing of energy of
$^{8}$Be.

\section{Acknowledgment}

This work was supported in part by the Program of Fundamental Research of the
Physics and Astronomy Department of the National Academy of Sciences of
Ukraine (Project No. 0117U000239) and by the Ministry of Education and Science
of the Republic of Kazakhstan, Research Grant IRN: AP 05132476.

\bibstyle{ieeetr}

\end{document}